\documentclass[aps,prl,twocolumn,superscriptaddress,10pt]{revtex4-2}

\usepackage{amsfonts}
\usepackage{mathtools}
\usepackage[svgnames,table]{xcolor}
\usepackage[T1]{fontenc}
\usepackage{txfonts}
\usepackage{graphicx}
\usepackage{lineno}
\usepackage{gensymb}
\usepackage{subcaption}

\usepackage{hyperref}
\hypersetup{
        colorlinks=true,
        citecolor=SteelBlue,
        filecolor=LimeGreen,
        linkcolor=SlateBlue,
        urlcolor=MediumPurple
}

\usepackage{dcolumn}
\usepackage{bm}
\usepackage{amsmath}
\usepackage{breqn}
\usepackage[bbgreekl]{mathbbol}
\usepackage{bbm}
\usepackage{mathrsfs}
\usepackage[permil]{overpic}
\usepackage{siunitx}

\usepackage{tikz}
\usetikzlibrary{calc} 

\DeclareSIUnit\sqrthz{\ensuremath{\sqrt{\text{Hz}}}}
\DeclareSIUnit[per-mode = symbol]\Hzasd{\Hz\per\sqrthz}
\DeclareSIUnit[per-mode = symbol]\masd{\m\per\sqrthz}
\DeclareSIUnit[per-mode = symbol]\pmasd{\pico\m\per\sqrthz}

\makeatletter
\let\cat@comma@active\@empty
\makeatother

\begin{document}

\preprint{APS/123-QED}
\title{Extending Ground-Based Gravitational-Wave Sensitivity to 5 Hz}
\author{Amit Singh Ubhi}
\email{a.s.ubhi@bham.ac.uk}
\affiliation{Institute for Gravitational Wave Astronomy, School of Physics and Astronomy, University of Birmingham, Birmingham B15 2TT, United Kingdom}
\author{Lari Koponen}
\affiliation{Institute for Gravitational Wave Astronomy, School of Physics and Astronomy, University of Birmingham, Birmingham B15 2TT, United Kingdom}
\author{Jiri Smetana}
\affiliation{Institute for Gravitational Wave Astronomy, School of Physics and Astronomy, University of Birmingham, Birmingham B15 2TT, United Kingdom}

\author{Yulin Xia}
\affiliation{Department of Physics, Tsinghua University, Beijing 100084, China}
\author{Haixing Miao}
\affiliation{Department of Physics, Tsinghua University, Beijing 100084, China}

\author{Emilia Chick}
\affiliation{Institute for Gravitational Wave Astronomy, School of Physics and Astronomy, University of Birmingham, Birmingham B15 2TT, United Kingdom}
\author{John Bryant}
\affiliation{Institute for Gravitational Wave Astronomy, School of Physics and Astronomy, University of Birmingham, Birmingham B15 2TT, United Kingdom}
\author{Geraint Pratten}
\affiliation{Institute for Gravitational Wave Astronomy, School of Physics and Astronomy, University of Birmingham, Birmingham B15 2TT, United Kingdom}
\author{Teng Zhang}
\affiliation{Institute for Gravitational Wave Astronomy, School of Physics and Astronomy, University of Birmingham, Birmingham B15 2TT, United Kingdom}

\author{Richard Mittleman}
\affiliation{LIGO Laboratory, Massachusetts Institute of Technology, Cambridge, Massachusetts 02139, USA}
\author{Peter Fritschel}
\affiliation{LIGO Laboratory, Massachusetts Institute of Technology, Cambridge, Massachusetts 02139, USA}

\author{Alan V. Cumming}
\affiliation{Institute for Gravitational Wave Research, School of Physics and Astronomy, University of Glasgow, Glasgow G12 8QQ, United Kingdom}
\author{Giles Hammond}
\affiliation{Institute for Gravitational Wave Research, School of Physics and Astronomy, University of Glasgow, Glasgow G12 8QQ, United Kingdom}

\author{Denis Martynov}
\affiliation{Institute for Gravitational Wave Astronomy, School of Physics and Astronomy, University of Birmingham, Birmingham B15 2TT, United Kingdom}

\date{\today}

\begin{abstract}

Extending the sensitivity of terrestrial gravitational-wave detectors below 20\,Hz is a long-standing challenge, limited by ground motion and inertial sensing noise. In this letter, we demonstrate ultra-high-vacuum compatible inertial isolation and position sensing technologies that achieve active platform stabilization down to 10\,mHz. Our laser position sensors reach a sub-pm/$\sqrt{\rm Hz}$ sensitivity above 10\,mHz, independent of the input light polarization, representing a 100-fold improvement over the current LIGO position sensors. In addition, our inertial sensors provide at least a factor of 5 improvement in low-frequency sensitivity compared to state-of-the-art commercial seismometers. 
We integrate these technologies into a LIGO-like interferometer model and predict a low-frequency sensitivity improvement of up to an order of magnitude at 10\,Hz, with enhanced linearity and calibration stability. This extension increases the detection horizon for intermediate-mass black hole binaries of mass $10^3 M_\odot$ by a factor of 3.
Our results provide the first experimental demonstration of a practical pathway to sub-10\,Hz operation of terrestrial gravitational-wave detectors and establish key technologies for next-generation observatories such as Cosmic Explorer and Einstein Telescope.

\end{abstract}

\maketitle

\textit{Introduction.-} Intermediate mass black holes (IMBHs) with masses between $100\,M_{\odot}-10^5\,M_{\odot}$ occupy a critical but largely unexplored gap between stellar-mass and supermassive black holes~\cite{Greene_2020}. Only two direct measurements of low-mass IMBHs have been confirmed from gravitational wave (GW) events GW190521 \cite{GW190521} and GW231123 \cite{GW231123}, providing compelling evidence for the existence of such objects. Their detections challenge conventional models of stellar evolution and black hole formation \cite{StellarCollapse, HeirarchicalMergers, GW231123_Stellar}, where the formation pathways bridging stellar-mass and supermassive black holes are poorly understood. Detection and characterization of IMBHs are essential for constraining black hole growth channels, testing hierarchical mergers, and probing the evolution of the Universe’s most massive compact objects, including the upper mass gap ($50\,M_{\odot} - 130\,M_{\odot}$) \cite{MassGapClusters,MassGapPairInstability3G,MassGapNuclearBurning}. The emission frequency of GWs scales inversely with the total source mass. Therefore, improving low-frequency sensitivity in terrestrial detectors such as LIGO \cite{GW150914, GWTC-1, GWTC-2, GWTC-3, GWTC-4} is essential to detect IMBHs.

Beyond IMBH detection, improved low-frequency sensitivity also enhances the signal-to-noise ratio (SNR) for lower-mass binaries, extends the pre-merger observation time, and sharpens sky-localization of sources. These gains increase the precision of parameter estimation, enable earlier alerts for multi-messenger follow-up \cite{GW170817,GW170817multi}, and reduce degeneracies in inferred source properties, maximizing the scientific return of gravitational-wave observations.

\begin{figure}[t]
    \centering
    \begin{tikzpicture}
        \node[anchor=south west] (figa) at (0,0) {\includegraphics[width=0.95  \linewidth]{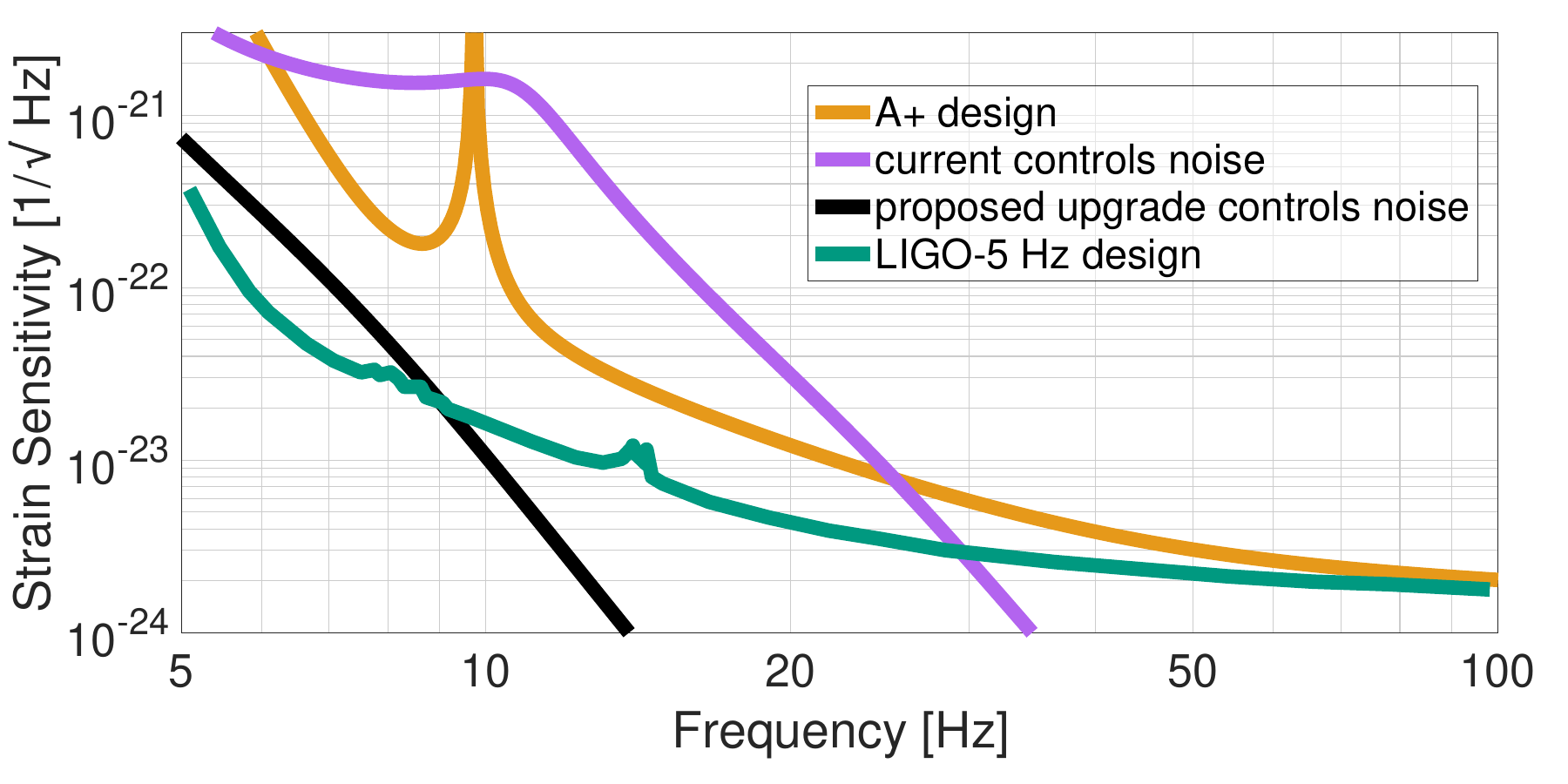}};
    \end{tikzpicture}
      \caption{Detector strain sensitivity comparison for the A+ design and LIGO-5\,Hz design~\cite{LIGO-LF}. The technical noise with the current inertial isolation schemes is shown in violet. The estimated improvement in the technical noises using our technologies (black) will enable the A+ design sensitivity to be reached, and demonstrates a clear pathway for terrestrial GW detection at 5\,Hz.}
    \label{fig:DHR}
\end{figure}

\begin{figure*}[t]
    \centering
    
\begin{tikzpicture}

        \node[anchor=center, inner sep=0] (figa) at (0,0)
            {\includegraphics[width=0.59\textwidth]{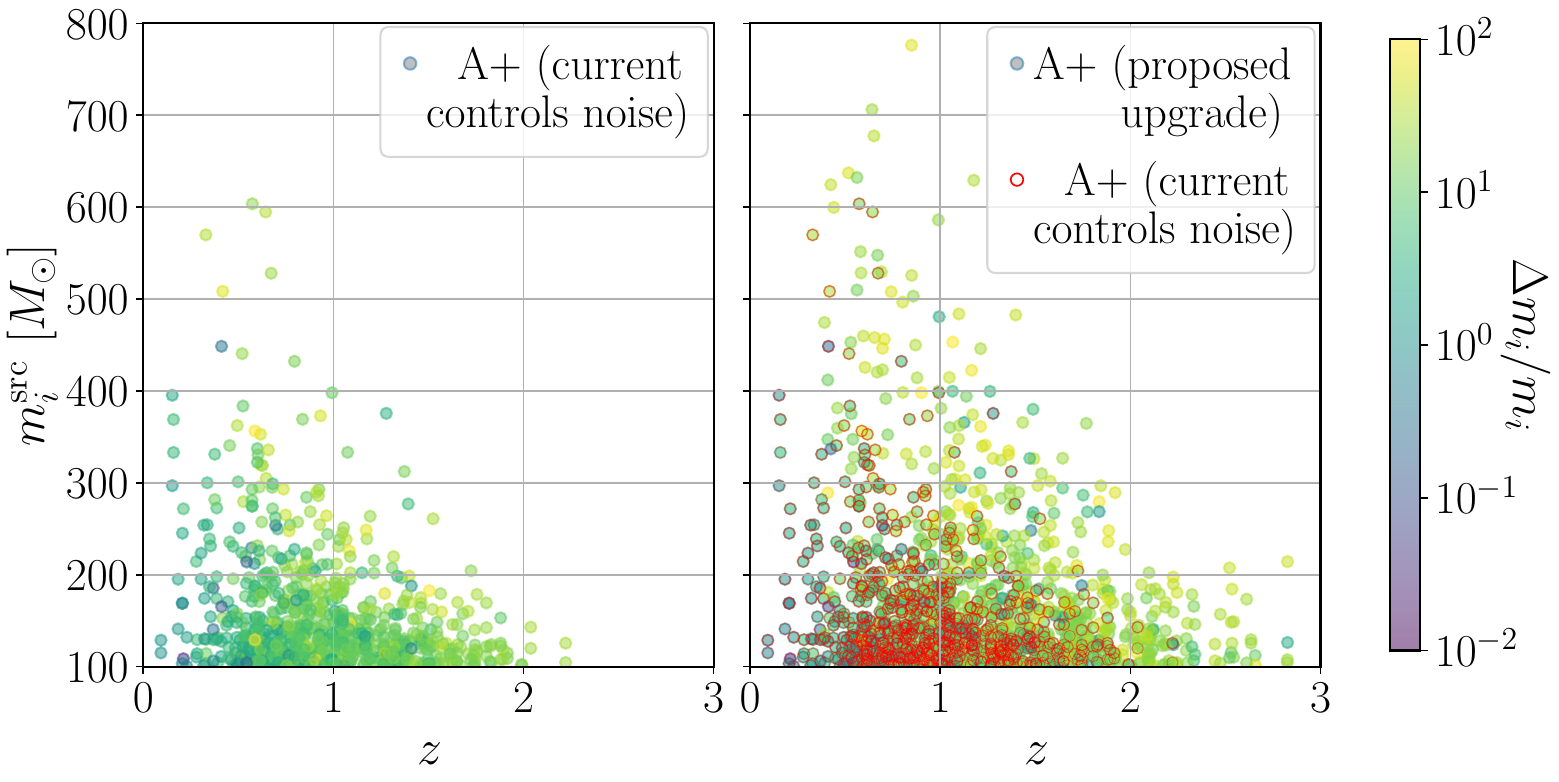}};

        \node[anchor=south west]
            at ($(figa.north west)+(0,0)$) {\textbf{(a)}};

        \node[anchor=west, inner sep=0] (figb)
            at ($(figa.east)+(-0,0)$)
            {\includegraphics[width=0.4\textwidth]{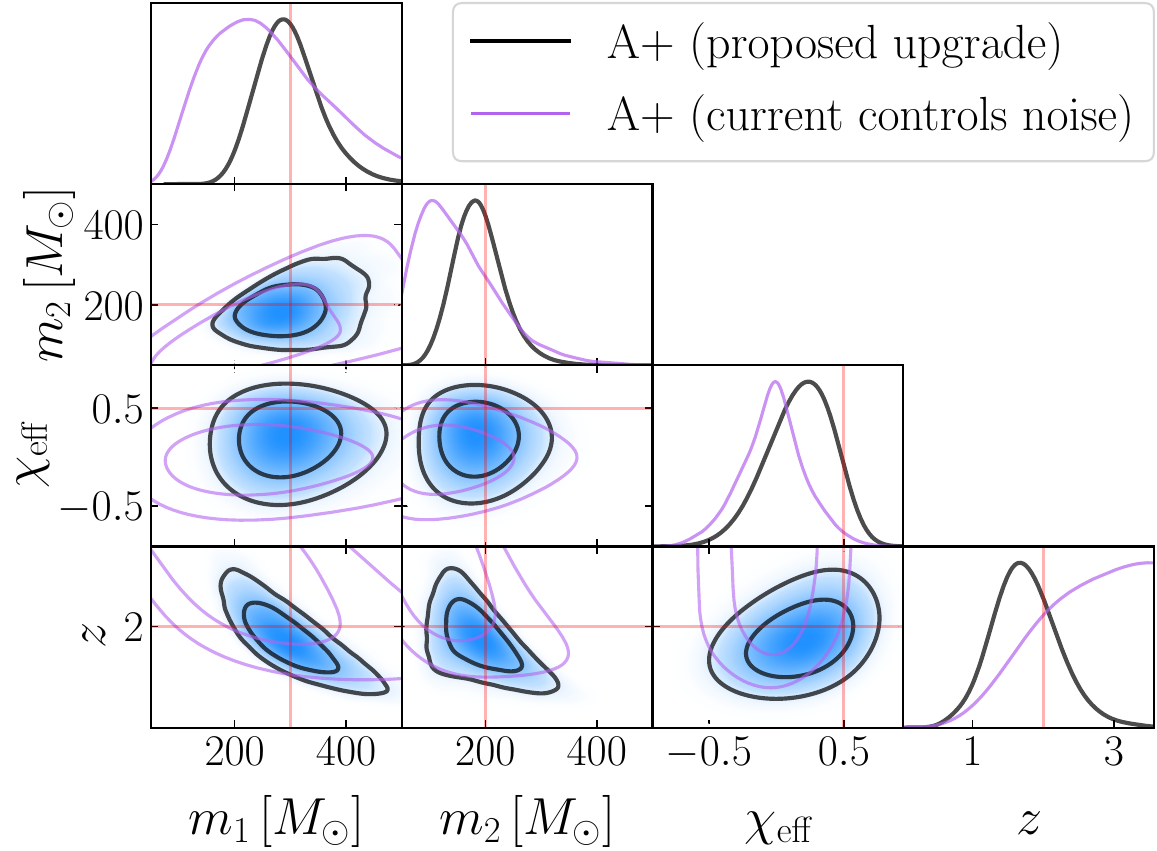}};

        \node[anchor=south west]
            at ($(figb.north west)+(0,0)$) {\textbf{(b)}};

    \end{tikzpicture}

    \caption{\textbf{(a)} Left: fractional error on the detector-frame masses as a function of redshift for the A+ configuration with the current controls noise. Right: comparison between the A+ configuration with the current control noise (red circles) and A+ design sensitivity when the proposed upgrade is implemented (colored circles). \textbf{(b)} Parameter estimation comparison for an IMBH merger with source-frame masses of $300\,M_{\odot}$ and $200\,M_{\odot}$ and dimensionless spin magnitudes of 0.9 and 0.8 respectively. The binary is placed at a redshift of $2$, near the horizon of A+ detector with the developed technologies.}
    
    \label{fig:param_est}
\end{figure*}

In this letter, we demonstrate three technologies that together will reduce the residual motion of all suspended optics in the detectors via reduced input motion and sensing noise. These technologies, combined, are critical for future upgrades \cite{LIGO-LF,ASharpCE} and third-generation detectors \cite{ASharpCE,CEphysics,CEsensitivity,ETscience,ETHild} due to stringent requirements for low-frequency seismic isolation \cite{MatichardReview, MatichardPart1, MatichardPart2} and suspension controls \cite{BOSEMcarbone, BOSEMsam}. We simulate improved strain sensitivity \cite{O1sensitivity,O3sensitivity,O4sensitivity,NoiseSubtraction} from their integration below 30\,Hz (Fig.\,\ref{fig:DHR}), achieving fundamental sensitivity limits in LIGO, and quantify their impact on IMBH detection and parameter estimation (Fig.\,\ref{fig:param_est}). 

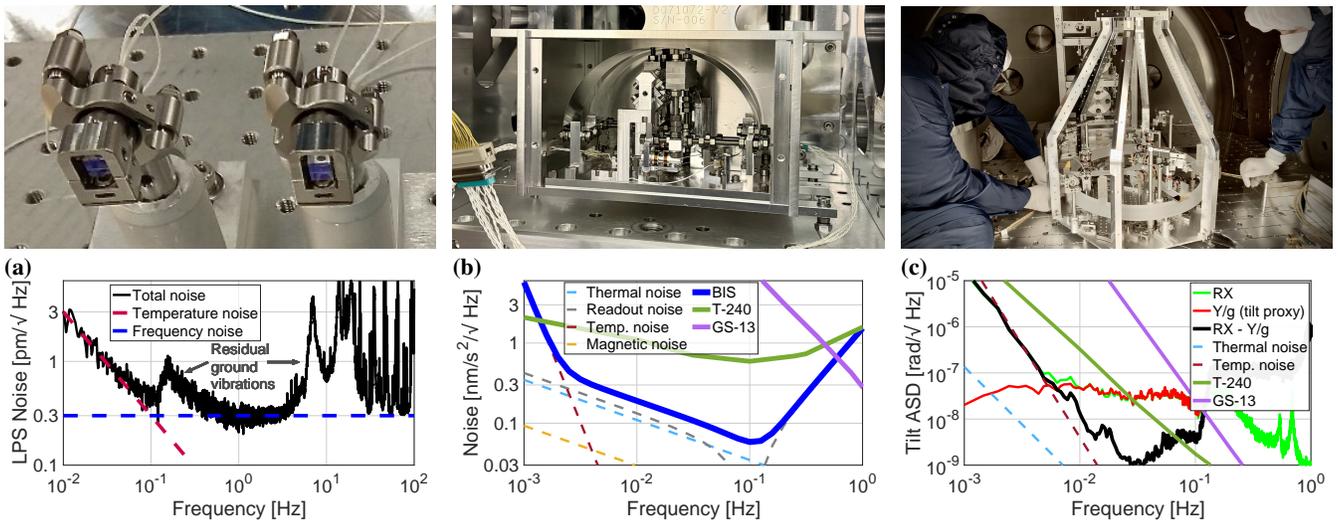
\begin{figure*}
 \centering
    \begin{tikzpicture}
        \node[anchor=south west] (figa1) at (0,0) {\includegraphics[width=0.32\textwidth]{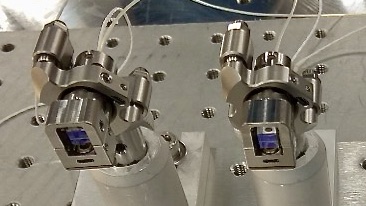}};

        \node[anchor=north west] (figa2) at (0,0)
        {\includegraphics[width=0.32\textwidth]{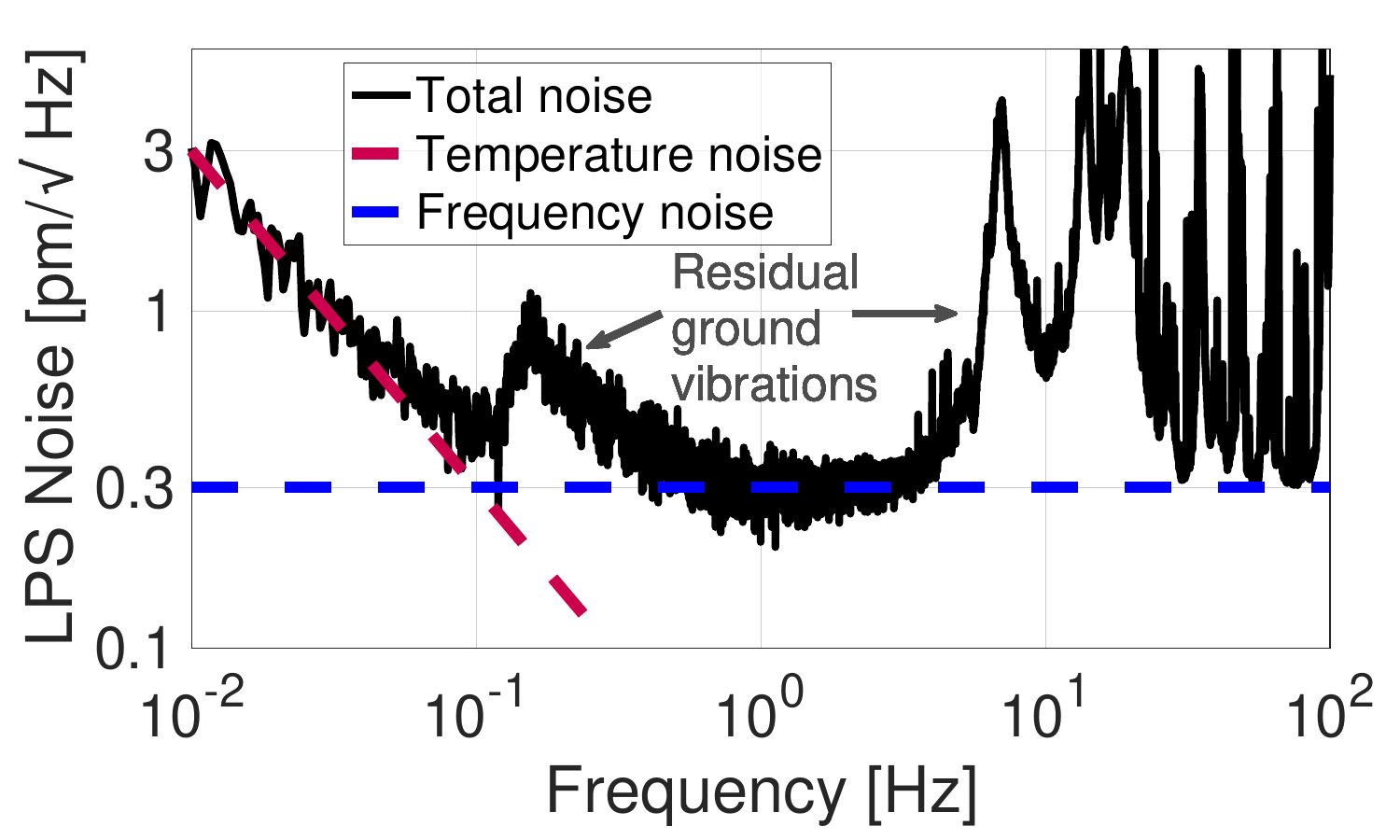}};
        \node[anchor=north west] at ($(figa2.north west) + (0,0.1)$){\textbf{(a)}};

        \node[anchor=south west] (figb1) at ($(figa1.south east) + (0,0)$) {\includegraphics[width=0.32\textwidth]{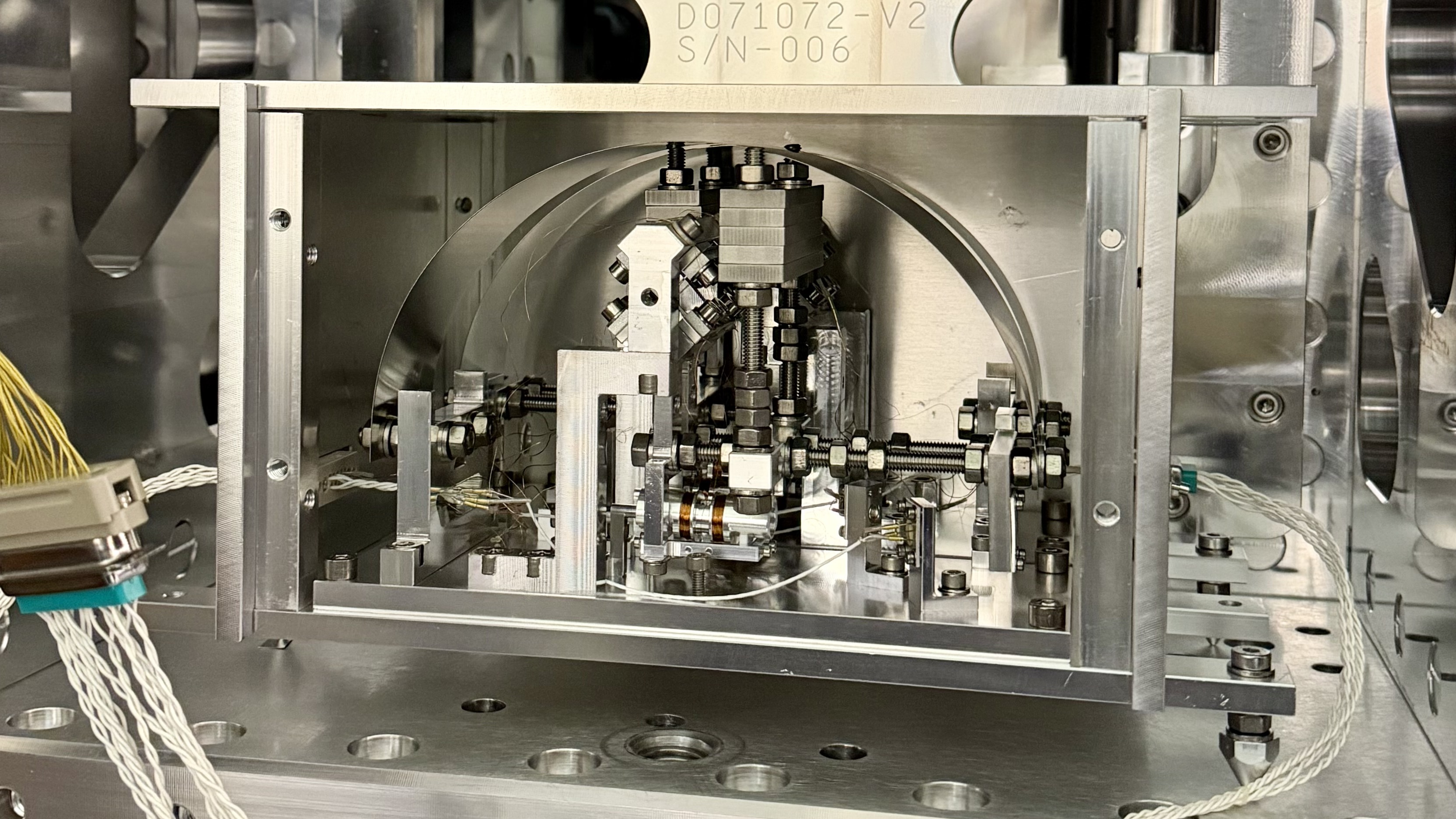}};

        \node[anchor=south west] (figb2) at ($(figa2.south east) + (0,0)$) {\includegraphics[width=0.32\textwidth]{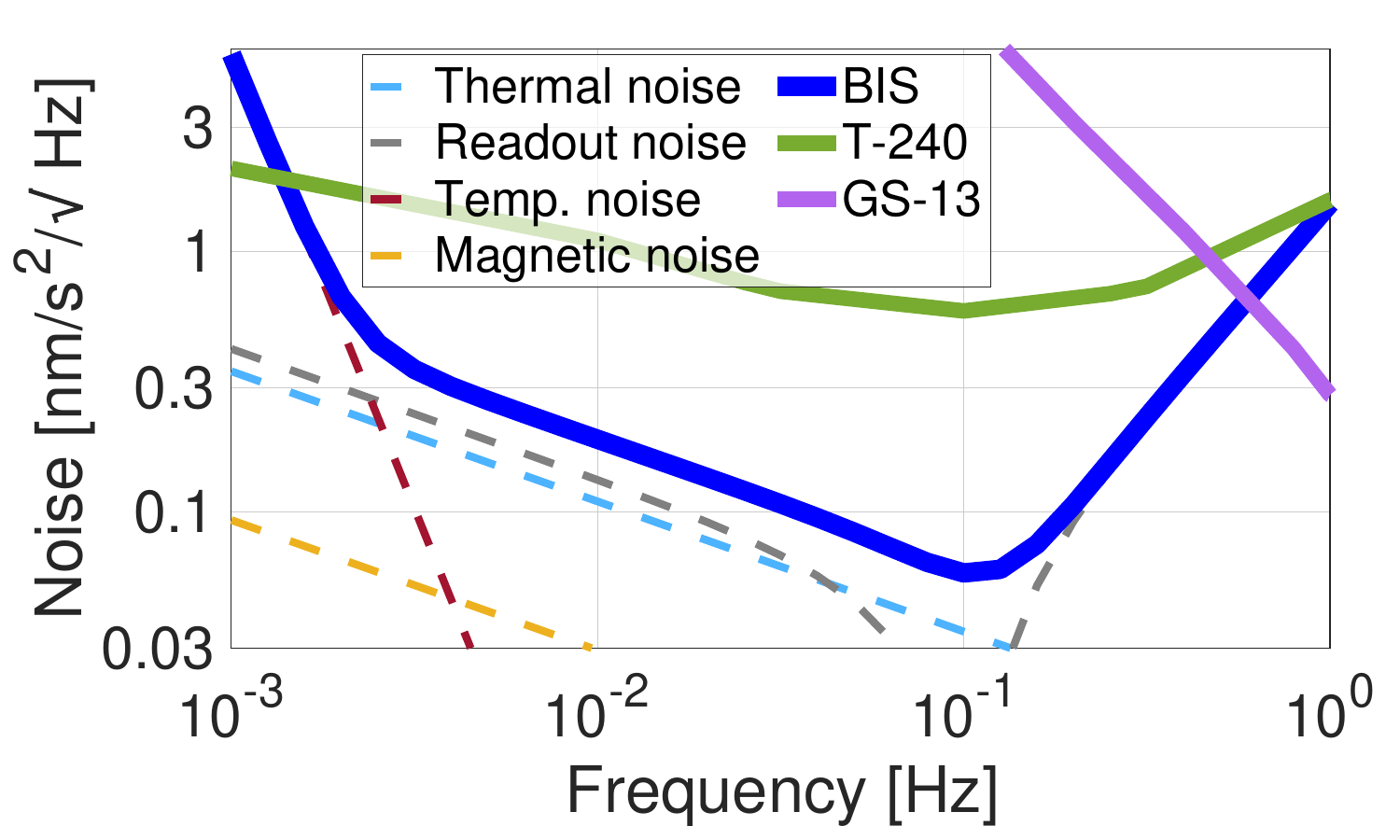}};
        \node[anchor=north west] at ($(figb2.north west) +(0,0.1)$){\textbf{(b)}};

        \node[anchor=south west] (figc1) at ($(figb1.south east) + (0,0)$) {\includegraphics[width=0.32\textwidth]{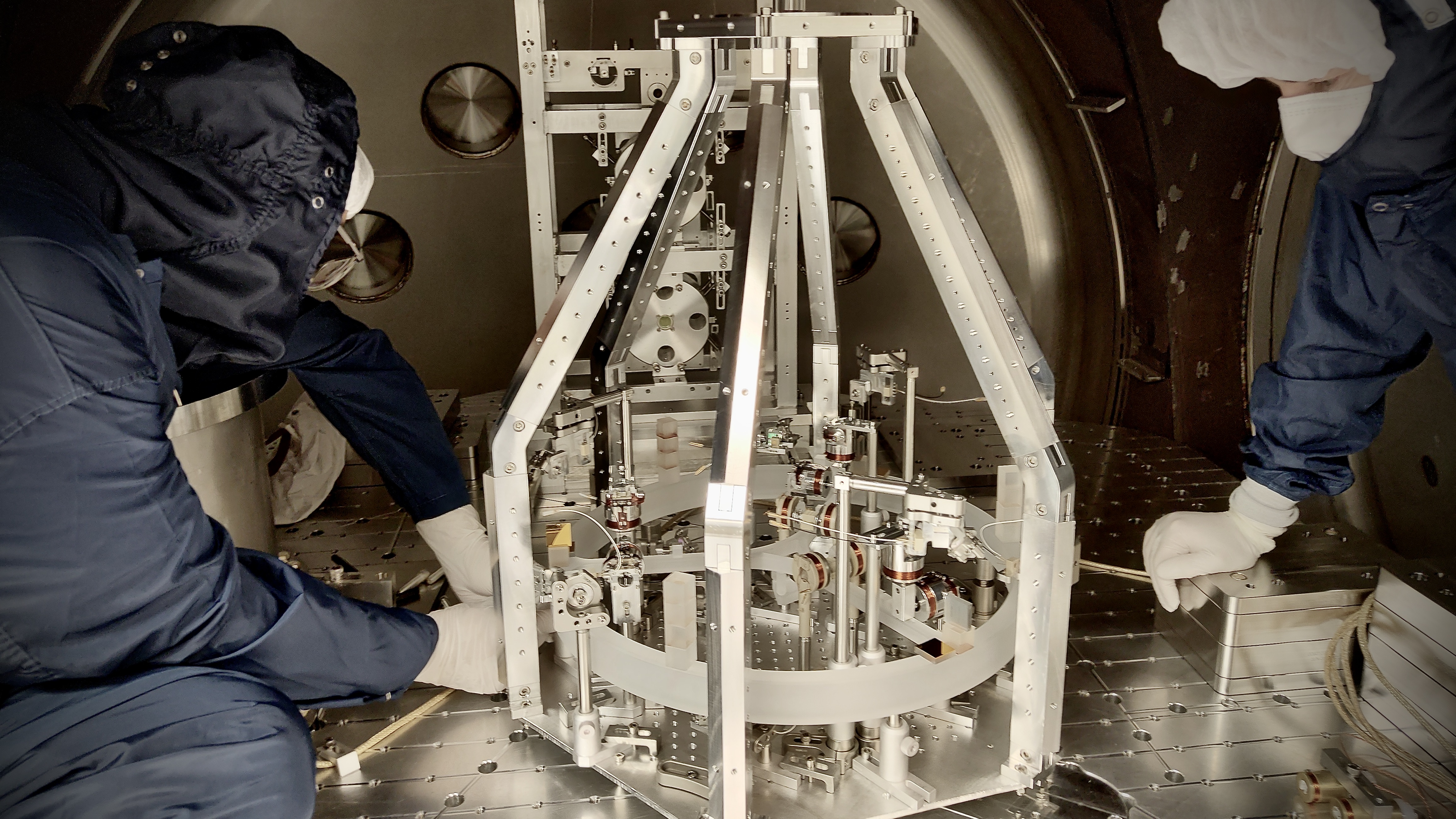}};

        \node[anchor=south west] (figc2) at ($(figb2.south east) + (0,0)$) {\includegraphics[width=0.32\textwidth]{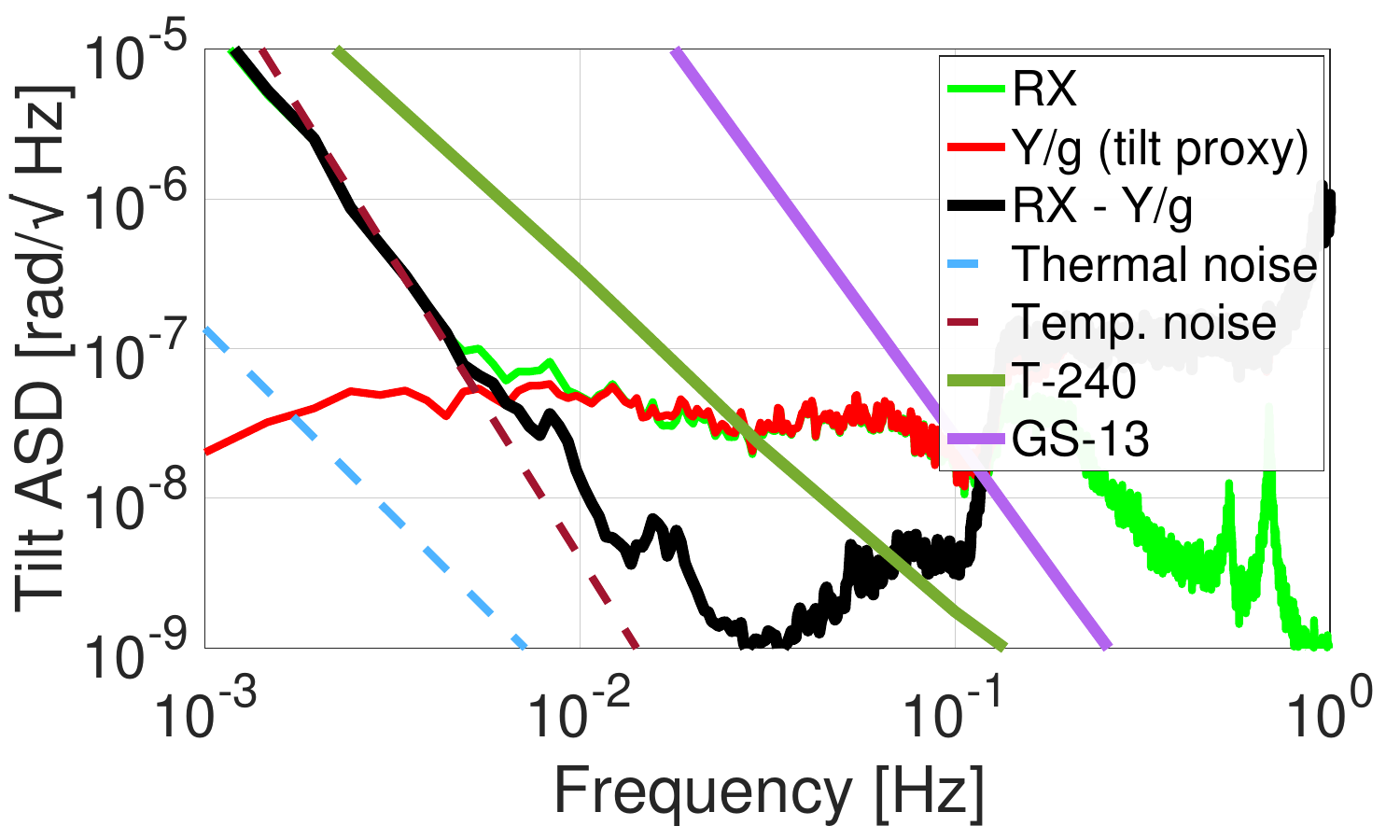}};
        \node[anchor=north west] at ($(figc2.north west) + (0,0.1)$) {\textbf{(c)}};
    \end{tikzpicture}
    \caption{%
    Photos of the three technologies and their respective noise budgets below.
    \textbf{(a)} Measured noise for the LPS (black). Subtraction of two sensors enables common-mode rejection of the DC-coupled laser frequency noise in the difference channel. Below 100\,mHz the noise is dominated by temperature couplings (dark red), whereas the high-frequency noise floor is dominated by down conversions of laser frequency noise from kHz demodulation (blue). Residual seismic couplings are present at the microseism and above 7\,Hz.
    \textbf{(b)} Noise budget for the BIS. Low frequency sensitivity is limited by direct temperature couplings. In the mid-frequencies, readout and thermal noises limit the device, with readout noise dominating above 0.1\,Hz. The total noise is shown in solid blue. The T-240 (dark green) and GS-13 (violet) noises are shown for reference.
    \textbf{(c)} Noise budget for the tilt degree of freedom for C-6D. Classical temperature noise limits the sensitivity below 10\,mHz. Above 30\,mHz we sit on the horizontal motion used to subtract the low frequency tilt motion (RX - Y/g) to access the noise floor (black). The T-240 (dark green) and GS-13 (violet) noise are shown for reference.
  }
  \label{fig:NB}
\end{figure*}

\textit{Technologies.-} The LIGO detectors use a combination of active and passive inertial isolation systems to minimize motion of their suspended optics~\cite{MatichardPart1,MatichardPart2,MatichardReview,LIGOsuspensions}. Active stabilization is achieved by controlling in-vacuum platforms that support the test masses and other core optics, including the beam splitter, and power- and signal-recycling cavity mirrors. The sensitivity of the inertial sensors on the platform determines its residual motion. Furthermore, active damping of the suspension resonances~\cite{LIGOsuspensions,BOSEMcarbone,BOSEMsam} is required to reduce their RMS motion. IMBH detection requires a reduction in the sensing noise of both the active platforms and the suspensions~\cite{LIGO-LF}. 
Recent improvements in the $3-30$\,Hz band have been demonstrated though optimization of control loop shapes using reinforcement learning methods~\cite{DeepLoopShaping}. In contrast, we focus on hardware-based sensing upgrades that reduce the fundamental sensing noise. These upgrades extend low-frequency performance toward the requirements of next-generation detectors.

The first technology is the compact interferometric Laser Position Sensor (LPS) \cite{SmarAct}, designed to replace the optical shadow sensors \cite{BOSEMcarbone, BOSEMsam} currently used for local damping of the suspension systems.

The LPS employs a standard telecom distributed feedback laser and deep frequency modulation interferometry \cite{DFMinterferometry,DPMinterferometry} in an unequal-arm Michelson configuration. The interferometric signal is encoded at the modulation frequency and its harmonics, enabling AC readout that shifts the measurement band away from low-frequency technical noises, and requires only one measurement channel for fringe counting. This sensor achieves a broadband sensitivity of $3 \times 10^{-13}\,\text{m}/\sqrt{\text{Hz}}$ (Fig.\,\ref{fig:NB}(a)) - an improvement of $\approx 100$ compared to the optical shadow sensors.

The key novelty of our design is the pentaprism beam splitter, which completely contains the reference arm of the interferometer.
The reference and sample beams then interfere with an offset from the input beam. This decouples reflected light from the optical fiber and suppresses non-linearities and excess noise caused by parasitic interference~\cite{smetana_2023_NL}.

Furthermore, the LPS is designed to be intrinsically insensitive to input polarization. In fiber-coupled systems, polarization rotation is unavoidable on the time scale of one year due to temperature fluctuations and stress relaxations even with the use of polarization-maintaining fibers.
We mitigate the effect of polarization drifts by matching the optical paths of the S- and P-polarization states.

Beyond the intrinsic optical robustness of the LPS design, deployment to measure the six degrees of freedom (DoF) of a suspended mass enables a sensing geometry in which common low-frequency laser frequency noise is subtracted in four of the six DoFs. This removes the need for additional reference sensors and auxiliary control loops, reducing system complexity and easing integration into the detectors.

The next two technologies in our inertial isolation package are advanced inertial sensors that suppress low-frequency motion. Commercial state-of-the-art seismometers do not provide the low-frequency sensitivity at the microseism to enable the platform control sufficient for IMBH astrophysics. We begin with the Birmingham Inertial Sensor (BIS), which is designed to integrate into the active platforms from which the input, output, power-recycling, and signal-recycling optics are suspended. The BIS is more than an order of magnitude more sensitive below 0.1\,Hz than the Geotech Instruments GS-13 sensor currently used on these platforms. To the best of our knowledge, it is also the world's most sensitive single-axis seismometer in the critical 10\,mHz - 100\,mHz frequency band, surpassing the industry-leading Nanometrics Trillium 240 (T-240) seismometer by a factor of 5.

The design is based on the Wielandt–Streckeisen approach \cite{LeafSpring}, employing leaf springs whose shear force provides a tunable anti-spring. In addition to sensitivity, a key advantage of our sensor is its ultra-high-vacuum compatibility. In-vacuum operation suppresses the limiting thermal noise caused by the readout capacitor plates in state-of-the-art seismometers and simplifies the integration into the LIGO chambers. The BISs are tuned to have a long period of $\approx 15$\,s, improving the low-frequency readout noise. We position the sensors in pairs inside aluminum and mu-metal shields to suppress temperature fluctuations and reject external magnetic fields. A photo and noise budget for the BIS is shown in Fig.\,\ref{fig:NB}(b). Three pairs of BISs enable low-frequency readout and control of six DoFs of the active platforms.

The final sensor is the fused silica six-degree-of-freedom (six-axis) seismometer (C-6D) \cite{6Dproposal,6Dproto,6Dplatform,6Dmetal,6Dsilica} developed for stabilization of LIGO's test masses and beamsplitter. 
The tilt sensitivity of our sensor is $\leq 10^{-8}\,\mathrm{rad}/\sqrt{\mathrm{Hz}}$ at 10\,mHz, a factor of 30 better than the state-of-the-art T-240 seismometer. The tilt noise budget is shown in Fig.\,\ref{fig:NB}(c). 

The proof mass of C-6D is a fused silica ring with a low index of thermal expansion ($<5\times 10^{-7}$/K) suspended by a fused silica fiber with a low mechanical loss angle ($<10^{-6}$)~\cite{SilicaFibre2012}. This proof mass has eigen frequencies of 1.5\,mHz, 9\,mHz, and 17\,mHz for the critical rotational DoFs, which are tuned to minimize the coupling of the sensing noise and simplify the platform control. The proof mass is sensed by six LPSs and its low-frequency drifts ($<1$\,mHz) are stabilized using coil-magnet actuators.


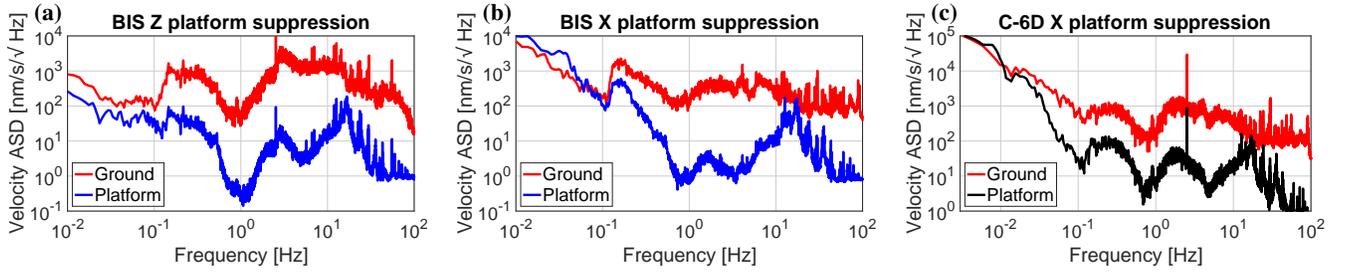
\begin{figure*}
 \centering
    \begin{tikzpicture}
        \node[anchor=south west] (figa) at (0,0) {\includegraphics[width=0.32\textwidth]{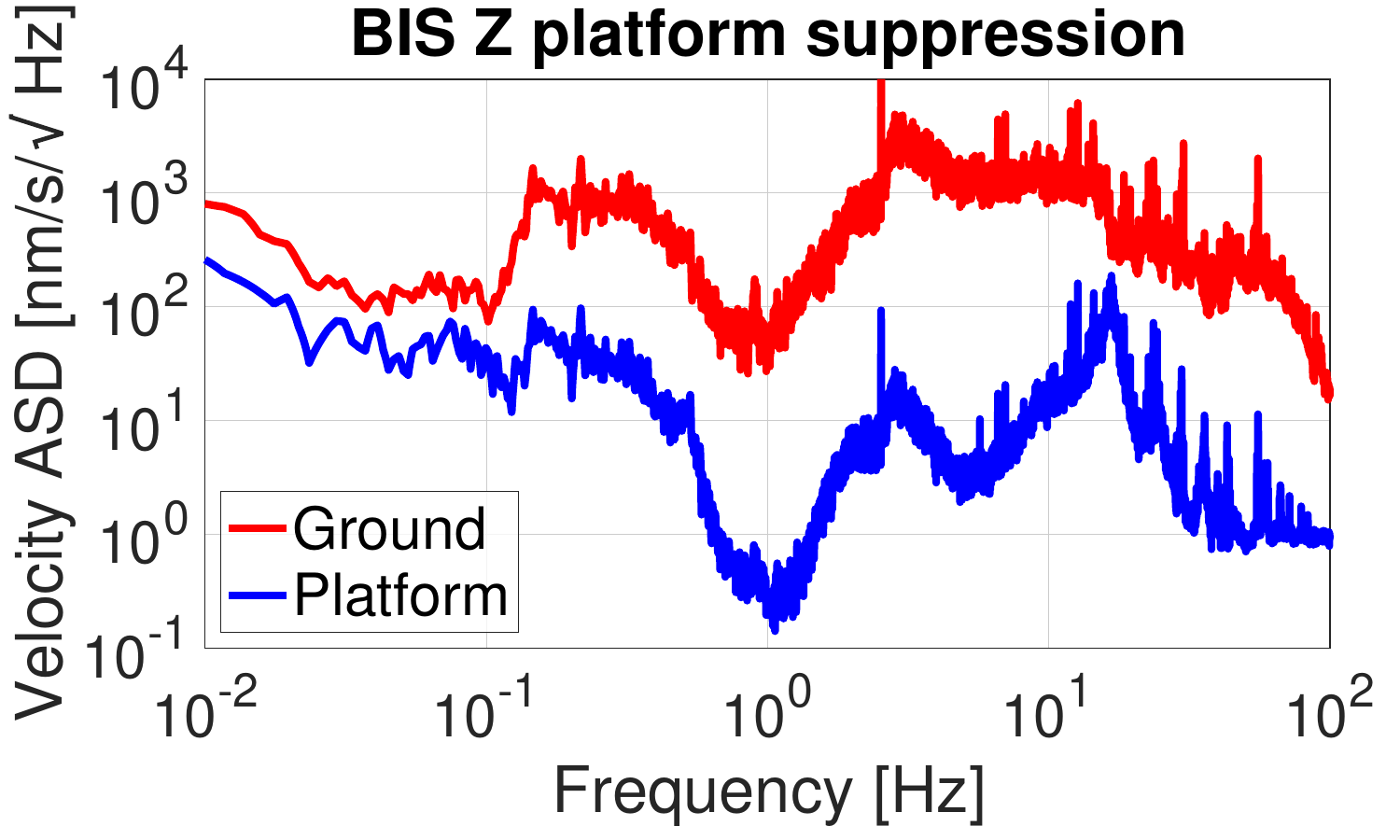}};
        \node[anchor=north west] at ($(figa.north west) +(0.4,0.1)$){\textbf{(a)}};

        \node[anchor=south west] (figb) at ($(figa.south east) + (0,0)$) {\includegraphics[width=0.32\textwidth]{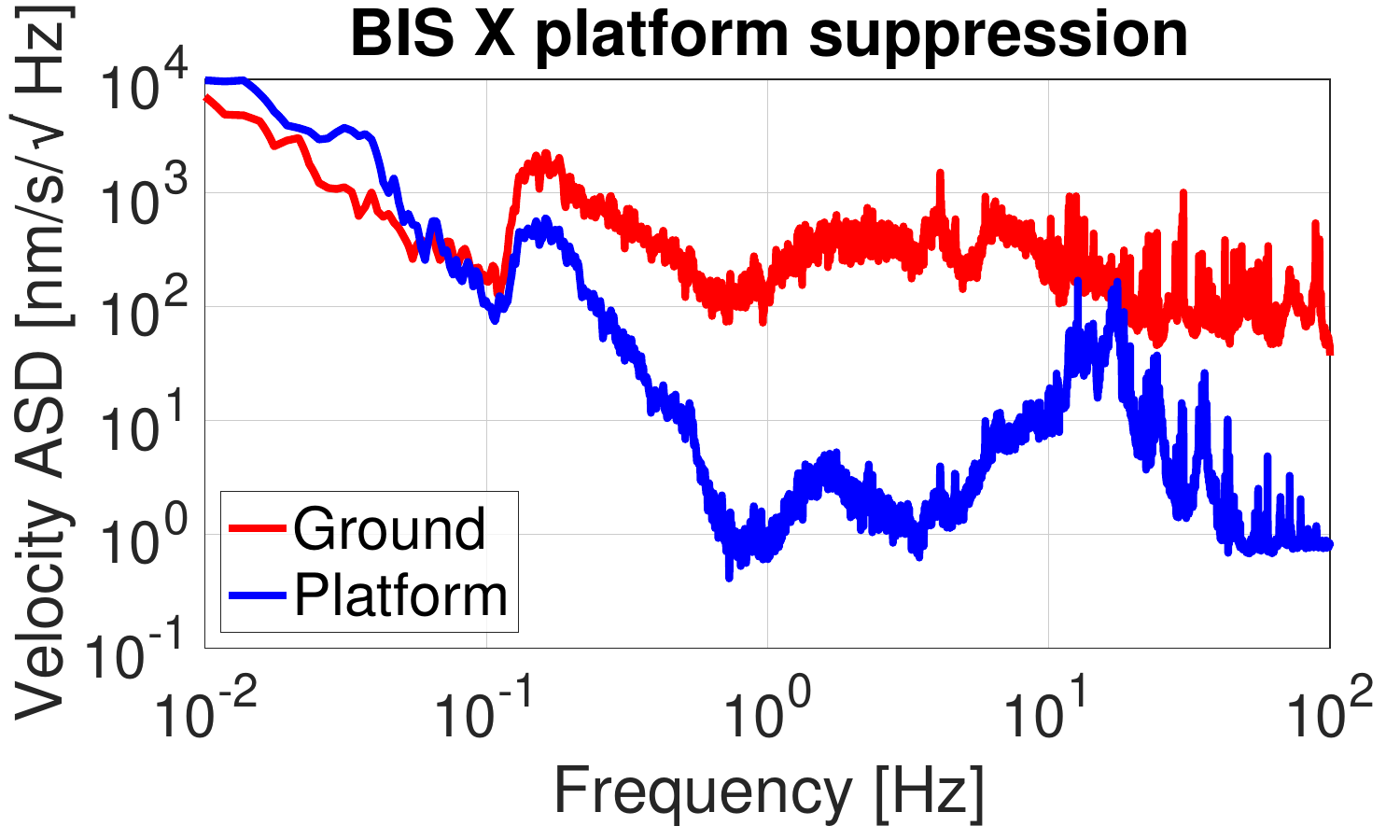}};
        \node[anchor=north west] at ($(figb.north west) +(0.4,0.1)$){\textbf{(b)}};

        \node[anchor=south west] (figc) at ($(figb.south east) + (0,0)$) {\includegraphics[width=0.32\textwidth]{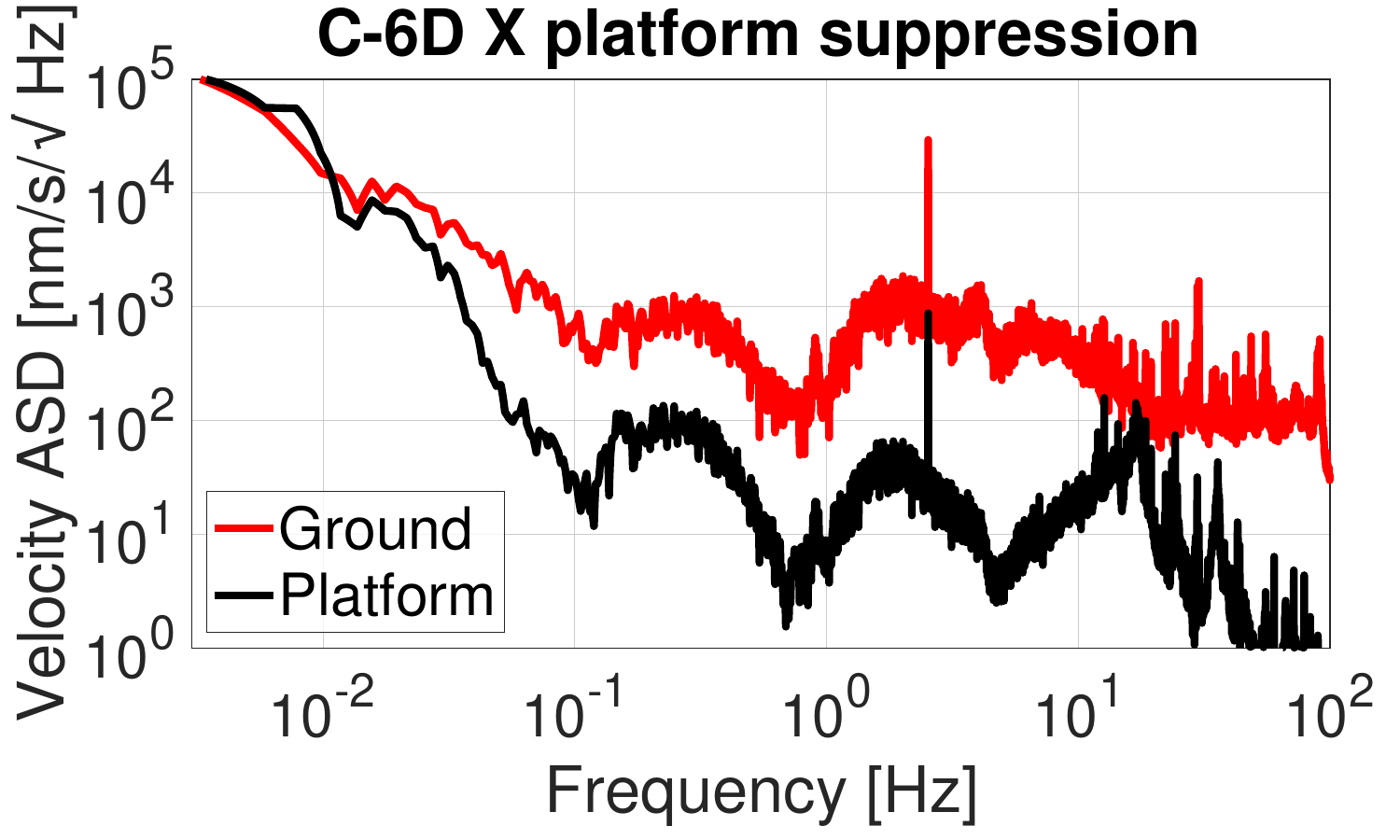}};
         \node[anchor=north west] at ($(figc.north west) +(0.4,0.1)$){\textbf{(c)}};

    \end{tikzpicture}
    \caption{%
        Active platform stabilization at LIGO's MIT facility. Red traces are the measured ground motion using the out-of-vacuum Streckeisen STS-2 seismometer. The blue (BISs in-loop) and black (C-6D in-loop) traces show the measured platform motion using a T-240 seismometer as a witness sensor.
        \textbf{(a)} Z degree of freedom.
        \textbf{(b)} X degree of freedom.
        \textbf{(c)} X degree of freedom.
    }
    \label{fig:MITsuppression}
\end{figure*}

\textit{Platform stabilization.-} Most active platforms suppress external vibrations from 0.1 - 1\,Hz and inject the sensor noise at lower frequencies~\cite{ActiveIsolation1991,MutliStageActiveIsolation,LowFVertIsolation,ActiveIsolation2020,Wang2025_OpticalTableIsolation}.
We demonstrated the performance of our sensors at LIGO's MIT facility, achieving inertial isolation in all six DoFs. The results show that the sensors are not only highly sensitive as discussed above, but can also be robustly diagonalized for platform sensing.
To the best of our knowledge, the sensors have achieved the world-leading level of inertial isolation to date.

The low-frequency longitudinal control of LIGO's active seismic isolation platforms is limited by tilt-to-horizontal coupling in the platforms' inertial sensors~\cite{Tilt,Tilt2,BRS2}. This coupling scales as $g/\omega^2$, degrading the isolation at the microseism and further injecting noise below 0.1\,Hz.
Both the BIS and C-6D surpass the tilt sensitivities of LIGO's current inertial sensors as we have shown in Fig.\,\ref{fig:NB}(b) and (c). This improvement directly translates into enhanced low-frequency longitudinal suppression, with the BIS reducing microseismic motion by a factor of 20 in the vertical DoF and a factor of 4 in the horizontal DoFs. In the case of C-6D, we demonstrate the first ever longitudinal platform suppression down to 10\,mHz, as shown in Fig.\,\ref{fig:MITsuppression}(c).

\begin{figure*}[t]
    \centering
    \begin{tikzpicture}
        \node[anchor=south west] (figa) at (0,0) {\includegraphics[width=0.5\textwidth]{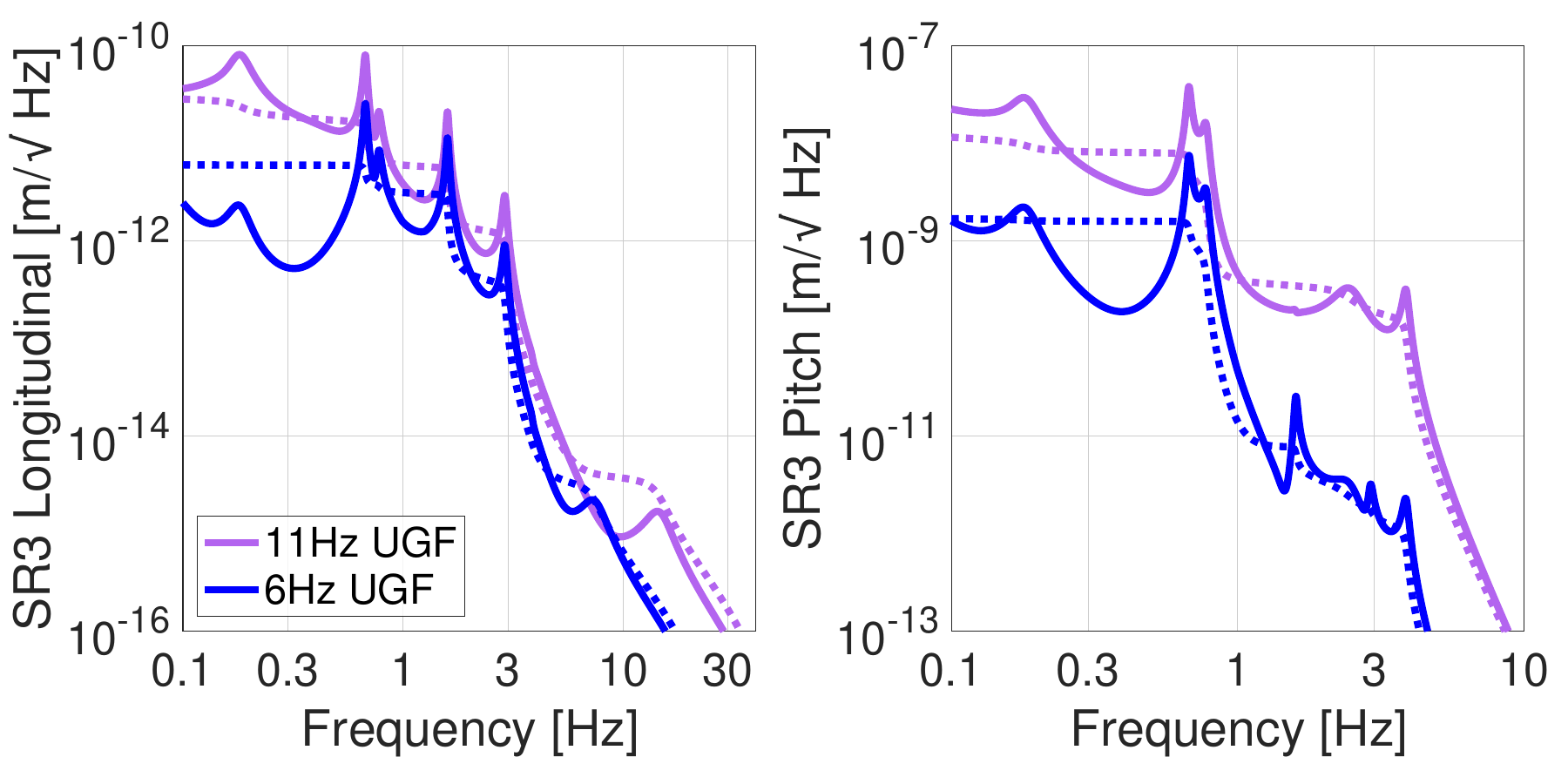}};
        \node[anchor=north west] at ($(figa.north west) + (0,0.1)$) {\textbf{(a)}};
        \node[anchor=north west] at ($(figa.north) + (-0.2,0.1)$) {\textbf{(b)}};

        \node[anchor=south west] (figb) at ($(figa.south east) + (-0.34,0)$) {\includegraphics[width=0.5\textwidth]{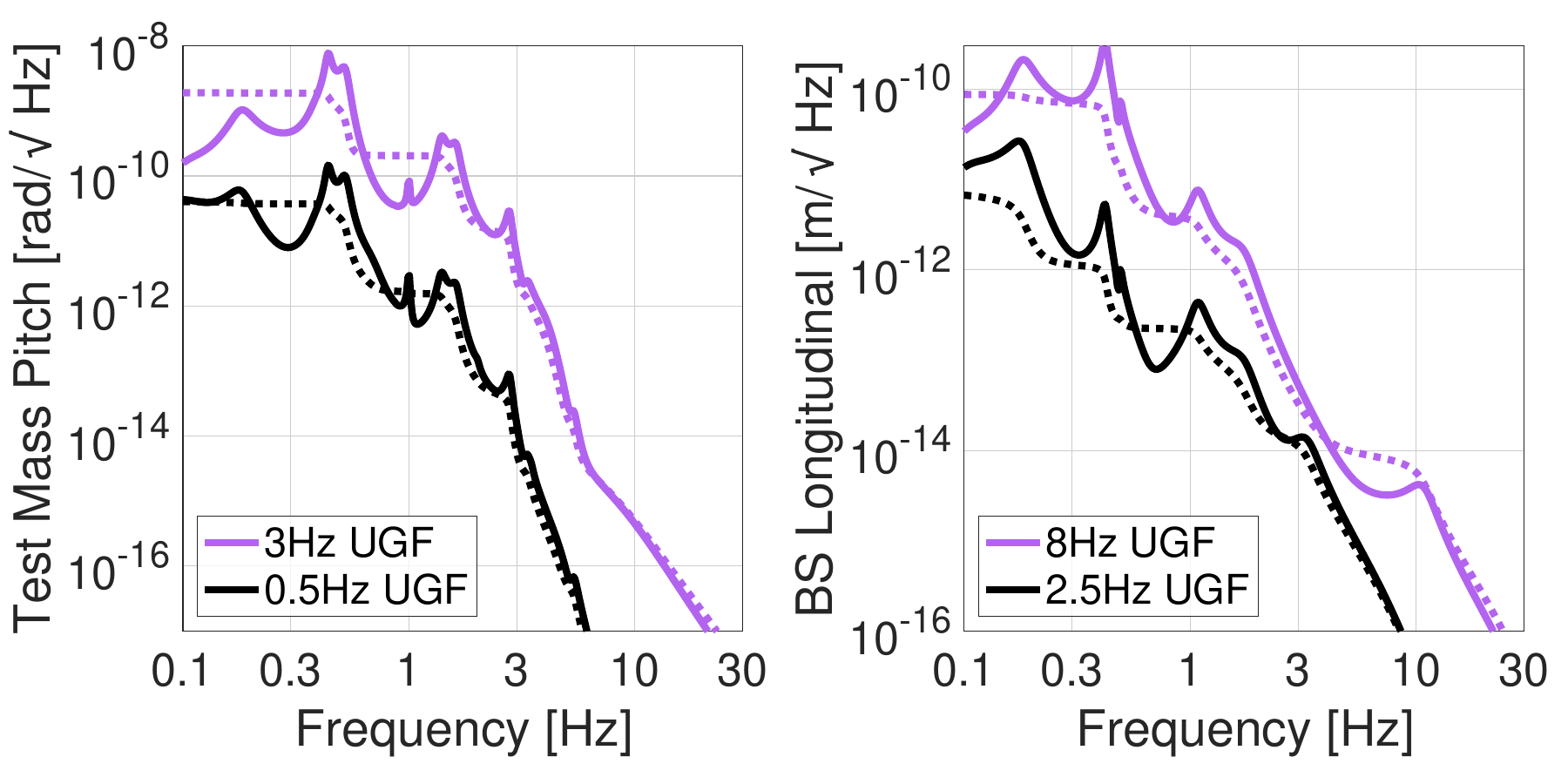}};
        \node[anchor=north west] at ($(figb.north west) + (0,0.1)$) {\textbf{(c)}};
        \node[anchor=north west] at ($(figb.north) + (-0.2,0.1)$) {\textbf{(d)}};
    \end{tikzpicture}
    \caption{%
        Comparison of different simulated optic motion when using the BISs (blue) and C-6D (black) in-loop to stabilize the relevant active platforms, and LPSs for suspension damping. The violet traces represent the current motion using the current control scheme with commercial seismometers and optical shadow sensors. The dotted lines represent the RMS motion.
        \textbf{(a)} Signal-recycling cavity telescope optic (SR3) longitudinal motion.
        \textbf{(b)} SR3 pitch motion.
        \textbf{(c)} Test mass pitch motion for an individual optic.
        \textbf{(d)} Beamsplitter longitudinal motion.
    }
    \label{fig:Optics}
\end{figure*}
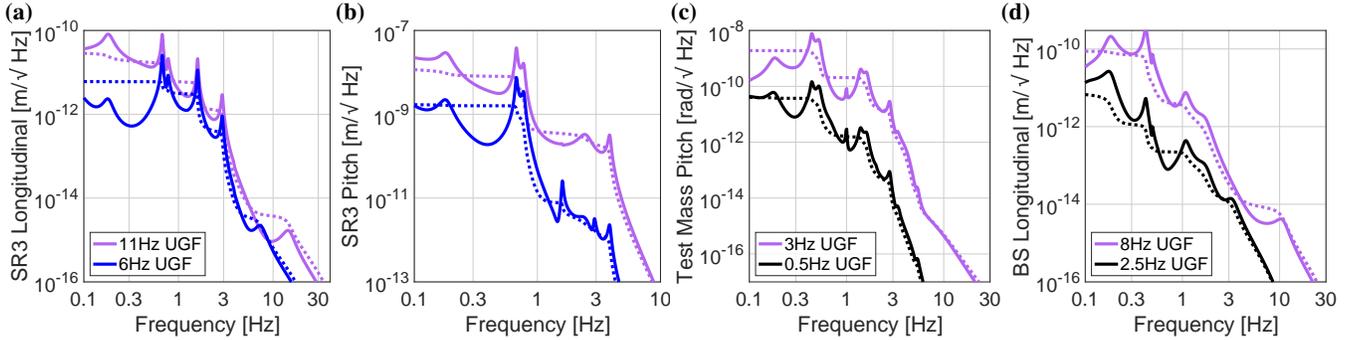

\textit{Detector improvements.-} The technologies have the potential to improve the sensitivity, linearity, and calibration of LIGO and next-generation GW detectors. We estimate these improvements by simulating the motion of the active platforms when using BIS and C-6D, and propagating the motion through the suspension chains of key optics, which are stabilized using LPSs. The resulting detector sensitivity is shown in Fig.\,\ref{fig:DHR}, and the astrophysical gains are illustrated via improved detection rate, range, source mass, and IMBH parameter estimation (Fig.\,\ref{fig:param_est}).

Two key control noises dominate the GW strain sensitivity below 30\,Hz \cite{O4sensitivity,O4characterisation}. First, angular sensing and control must stabilize the motion of the arm cavity mirrors to maintain a stable cavity axis. This is necessary due to the large optical power which can lead to radiation-pressure-induced angular instabilities~\cite{SidlesSigg,Liu2018_AngularInstabilityCavity}.
The second dominant noise arises from length sensing and control, primarily due to longitudinal beamsplitter motion. Feedforward to the test masses \cite{MICHfeedforward} partially mitigates this, however, the high bandwidth of the control loops leads to noise injection into the GW detection band. Integration of C-6D and LPSs reduces the control bandwidths from 3\,Hz to 0.5\,Hz for test mass pitch, and 8\,Hz to 2.5\,Hz for Michelson length control, while simultaneously reducing the RMS motion of the optics by a factor of 10 (Fig.\,\ref{fig:Optics}(c) and (d)). This suppresses control noise injection below the detector's low-frequency noise floor.

Next, we examine the power- and signal-recycling cavities, which directly impacts the power build up in the 4-km long cavities, the optical gain of the GW signal, and the detector linearity \cite{SRCopticalspring, RecyclingPI}. Both recycling cavities embed folded telescopes that reduce the beam size from $\approx 6$\,cm down to 1\,mm. Angular motion of the telescope mirrors changes the position and angle of the optical axes of the recycling cavities. This process couples noise, such as laser frequency and intensity noise, into the GW channel, modulates the antenna response, and compromises detector calibrations.
Implementation of BISs and LPSs will reduce the RMS longitudinal and pitch motion of the telescope optic by a factor of 5 (Fig.\,\ref{fig:Optics}(a) and (b)) and suppress output pointing fluctuations~\cite{O1sensitivity,MLmicroseism} by the same factor. This reduces parasitic coupling of the control sidebands' higher-order modes into the gravitational-wave channel and improves the detector linearity~\cite{Heinzel_recycling_mode_cleaning}.

With C-6D, BISs, and LPSs deployed on all platforms and suspensions, the residual motion of every suspended optic would be reduced in all DoFs. This would suppress low-frequency technical noises, including auxiliary length and alignment controls, suspension actuation and damping noise~\cite{BOSEMcarbone,BOSEMsam}, and scattered light~\cite{ScatteringVirgo,ScatteringO3,LIGO-LF}.
Reduced low-frequency platform motion would also minimize fringe wrapping and associated sensing nonlinearities~\cite{LIGOResonance}, further suppressing control-induced noise and enabling operation at the LIGO design sensitivity (Fig.~\ref{fig:DHR}).

\textit{Astrophysical gains.-} The low-frequency sensitivity improvements provide broad astrophysical gains across compact-binary systems. This includes enhanced detection rates, improved parameter estimation, and earlier inspiral access. We highlight the impact on IMBHs, for which the horizon and range increase substantially.
We quantify the improvement by using the Fisher information matrix to estimate the statistical uncertainties associated with the binary parameters following the analysis in~\cite{ASharpCE}, assuming a two-detector network consisting of LIGO Hanford and LIGO Livingston. 
The Fisher matrix is calculated using \texttt{GWFish}~\cite{Dupletsa:2022scg} and the \texttt{IMRPhenomXHM} waveform model~\cite{Pratten:2020fqn,Garcia-Quiros:2020qpx}. 
With few observational constraints, the functional form of the IMBH population is very poorly constrained and any figure of merit will depend on the assumed population. 
We generate a mock population assuming $1$ year of observation, with component masses drawn from a power-law distribution as in~\cite{ASharpCE}, isotropic spins with uniform magnitudes, and a merger rate following the Madau-Dickinson star formation rate~\cite{Madau:2014bja}, normalized to a local merger rate density of $1 \, \rm{Gpc}^{-3} \, \rm{yr}^{-1}$. 
We impose a detection threshold on the network SNR of $\rho_N \geq 12$. 
In Fig.~\ref{fig:param_est}(a), we show the fractional error on the detector frame masses as a function of redshift. 
The left panel shows the current $\textrm{A}+$ design with the present controls noise and the right panel the improvement that can be obtained when using the proposed upgraded configuration.  
For our assumed population, the upgraded configuration reduces the median Fisher uncertainty in the detector frame component masses by $\approx 50\%$ relative to the current controls scheme, and by around $\approx 64\%$ in the luminosity distance.  
We observe a relative increase of $\approx 86\%$ in the fraction of detected binaries in our population. In addition, the 90th percentile of the detected redshift distribution increases from $z_{90} \simeq 1.49$ to $z_{90} \simeq 1.84$, corresponding to an increase of $\approx 1.47$ in the enclosed comoving volume out to $z_{90}$.
The low-frequency improvements provided by the developed technologies offer a significant increase in the cosmological reach of the detectors for low-mass IMBHs and opens up a new discovery space for high-mass IMBH sources. 

To demonstrate the gain from the low-frequency improvements, we simulate an IMBH merger with component masses of $m_1 = 300\,M_{\odot}$ and $m_2 = 200\,M_{\odot}$, dimensionless spins of $\chi_1 = 0.9$ and $\chi_2 = 0.8$, and moderate spin-precession, $\chi_p = 0.5$~\cite{Schmidt:2014iyl}.
The binary is placed at a redshift of $z = 2$, sitting close to the horizon of the proposed detector as shown in Fig.~\ref{fig:DHR}.  
Assuming zero noise, the SNR is $\approx 9.2$ compared to $\approx 3.9$ in the presence of the current controls noise, where it falls below the detection threshold. 
To capture the full complexity of the signal near the merger regime, we use the NR surrogate model NRSur7dq4~\cite{Varma:2018mmi} for both injection and recovery. 
We performed Bayesian inference using \texttt{bilby}~\cite{Ashton:2018jfp} and the nested sampling package \texttt{dynesty}~\cite{Speagle_2020}. 
The joint posterior probability distributions for a subset of parameters is shown in Fig.\,\ref{fig:param_est}(b). 
Despite the challenging source being near the detection threshold, the proposed upgrade configuration resolves the total source frame mass to $M = 479^{+167}_{-112}$ and the redshift to $z = 1.7^{+0.9}_{-0.7}$. In contrast, the current controls noise provides no meaningful constraints, with the redshift being prior dominated. 
Such a clear observation of an IMBH would have profound implications for our understanding of black-hole formation channels, stellar evolution, and the growth of heavy black holes across cosmic time~\cite{Greene_2020}. 
This discovery potential is driven by the improvements in low-frequency detector sensitivity outlined here.  

Although we emphasize IMBHs as a compelling example, the low-frequency sensitivity improvements benefit compact-binary systems across the mass spectrum. Access to lower gravitational-wave frequencies extends the observable inspiral phase, since the time to merger scales as $t_{\rm merge} \propto f_{\rm GW}^{-8/3}$~\cite{Creighton2011_book}. Improved sensitivity below 30\,Hz therefore increases the accumulated SNR, enhances parameter estimation, and enables earlier pre-merger localization for loud and nearby events. These gains strengthen multi-messenger opportunities and improve cosmological and population studies more broadly. A comprehensive discussion of the scientific impact of low-frequency sensitivity is given in Refs.~\cite{LIGO-LF, CEphysics}.

\textit{Conclusions.-} In this letter, we present three complementary technologies enabling substantial improvements in LIGO’s low-frequency sensitivity to IMBHs. These technologies are required to approach the fundamental noise limits of current detectors and are critical for future upgrades and third-generation observatories. We characterize their noise performance and demonstrate their integration within the LIGO infrastructure, including active platform suppression tests at LIGO’s MIT facility. We also achieve longitudinal platform suppression down to 10\,mHz for the first time using C-6D. Propagating the impact of these technologies to detector strain sensitivity, we predict a $\approx 86\%$ increase in detectable IMBH binaries. Additionally, we simulate an IMBH merger at redshift 2, demonstrating parameter constraints not achievable with LIGO’s current isolation schemes.

Extending inertial isolation down to 10\,mHz fundamentally shifts stabilization from suppressing vibrations to actively controlling slow drift. This has the potential to further advance the LIGO sensitivity~\cite{LIGO-LF}, for example by increasing the inertia of the test masses and upgrading the suspension fibers~\cite{LIGO-G2301738}, and reach the low-frequency facility limit determined by the gravity-gradient noise~\cite{Hughes1998gravitygradient}.

Apart from the gravitational-wave detectors, the developed technologies have the potential to unlock major gains across multiple precision fields~\cite{Collette2010_vibration_isolation}: in semiconductor lithography it can reduce low-frequency stage motion by one to two orders of magnitude, and along with more sensitive displacement sensing~\cite{DisplacementInterferometry}, improve overlay accuracy and enable smaller feature sizes~\cite{PhotolithographyControl}; in optical telescopes it can suppress slow structural and thermal distortions, improving pointing and wavefront stability and enhancing high-contrast imaging~\cite{LeonGil2025_Telescope_isolation}; in quantum sensors and atomic interferometers it can reduce low-frequency acceleration noise, extending coherence times and enabling orders-of-magnitude sensitivity improvements~\cite{Xie2021_gravimeter_isolation,Oon2022_Atomic_gravimeter_isolation}; and in electron microscopy and precision metrology it suppresses drift that currently limits resolution and long-duration measurements~\cite{Shin2020_electron_microscope_isolation}. 

\section{Acknowledgements}
We acknowledge members of the LIGO Suspension
Working Group for useful discussions, the support of
the Institute for Gravitational Wave Astronomy at the
University of Birmingham, STFC Consolidated Grant
“Astrophysics at the University of Birmingham” (No.
ST/S000305/1), UKRI Quantum Technology for Fundamental Physics scheme
(Grant No. ST/T006609/1 and ST/W006375/1), and UKRI “The next-generation
gravitational-wave observatory network” project (Grant
No. ST/Y00423X/1). H.M. would
like to acknowledge the support from the National Key
R\&D Program of China “Gravitational Wave Detection”
(Grant No.: 2023YFC2205800), and support from Quantum Information Frontier
Science Center.
ASU thanks SKU and AKU for their encouragement and support.

\bibliographystyle{apsrev4-2}  
\bibliography{main.bib}

\end{document}


\title{Supplementary Material}

\maketitle

\section{Birmingham Inertial Sensor Noise}

\begin{figure}[h]
    \centering
    \includegraphics[width=\linewidth]{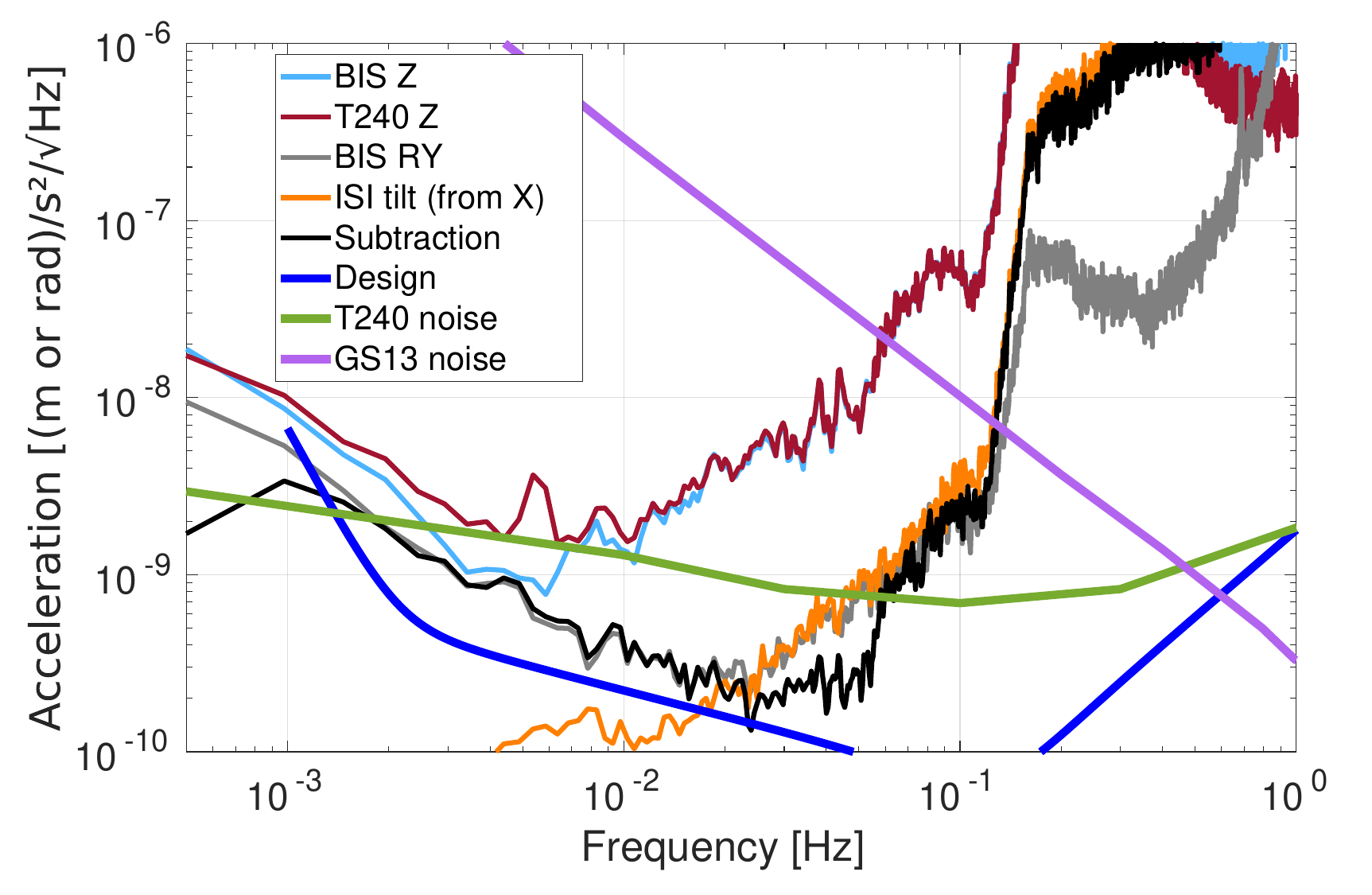}
    \caption{Measured platform tilt using the BIS (gray). The orange trace is the inferred platform tilt from the horizontal T-240 measurement. Subtraction of the inferred tilt (black) confirms that above 25\,mHz the BIS measures platform tilt motion.}
    \label{fig:BISNB}
\end{figure}

The Birmingham Inertial Sensors (BISs) are designed with ultra-high-vacuum compatibility for integration into LIGO's single stage active platforms. They are engineered to have improved noise compared to the Trillium T-240 in the $10-100$\,mHz frequency band, required for optimal active platform stabilization. During the MIT active platform control tests (discussed below), the BIS sensitivity was measured by analyzing the platform tilt (Fig.~\ref{fig:BISNB}). The BIS sensitivity surpassed the noise floor of the T-240 seismometer by a factor of 5 in the relevant frequency band. The orange trace shows the inferred tilt of the platform, measured by the $g/\omega^2$ tilt-to-horizontal coupling~\cite{Tilt,Tilt2} in the T-240 horizontal measurements. The BISs measured platform tilt motion above 25\,mHz (gray trace). Below 25\,mHz the device was noise limited. The T-240 self-noise and GS-13 self-noise are shown for comparison.

\section{MIT active platform control}

\begin{figure}[h]
    \centering
    \includegraphics[width=\columnwidth]{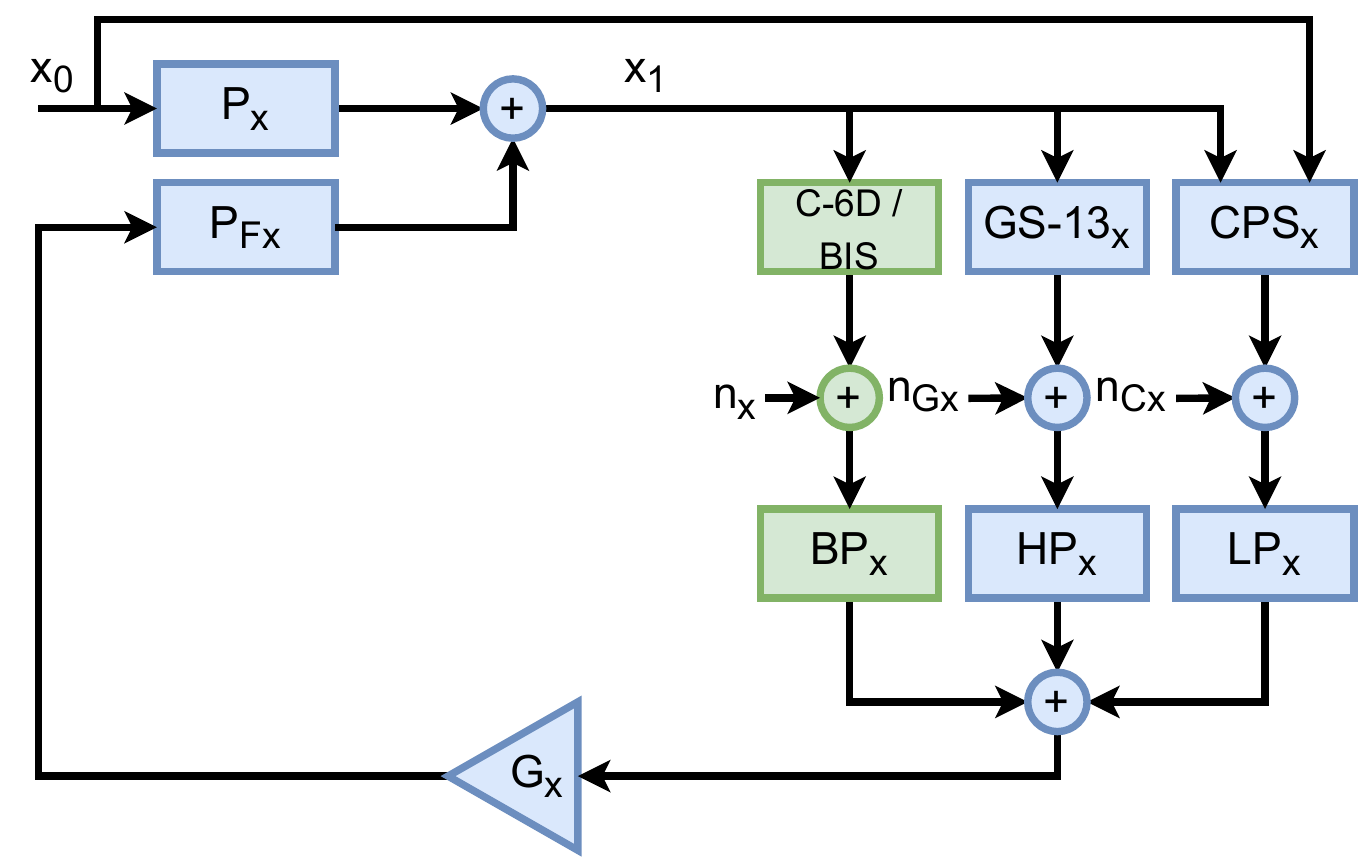}
    \caption{Single degree of freedom control loop for the active platform stabilization at LIGO's MIT facility. We integrate C-6D or BISs in the control path by bandpassing the sensing between the capacitive position sensors and the GS-13 inertial sensors which are currently used for stabilization.}
    \label{fig:MIT-loop}
\end{figure}

\begin{table}[h]
    \caption{Blending frequencies for the active platform stabilization tests at LIGO's MIT facility using the BISs.}
    \begin{tabular}{ccc}
    \hline
    \hline
     &  \multicolumn{2}{c}{Blend frequency} \\
    Degree of freedom & CPS to BIS & BIS to GS-13 \\
    \hline
    $X$ & 40\,mHz & 0.3\,Hz \\
    $Y$ & 80\,mHz & 0.3\,Hz \\
    $Z$ & 20\,mHz & 0.3\,Hz \\
    $RX$ & 67\,mHz & 0.3\,Hz \\
    $RY$ & 32\,mHz & 0.3\,Hz \\
    $RZ$ & 67\,mHz & 0.3\,Hz \\
    \hline
    \hline
    \end{tabular}
    \label{tab:BISblend}
\end{table}

\begin{table}[h]
    \caption{Blending frequencies for the active platform stabilization tests at LIGO's MIT facility using C-6D. Note we do not use C-6D for platform stabilization in the vertical degree of freedom.}
    \begin{tabular}{ccc}
    \hline
    \hline
     &  \multicolumn{2}{c}{Blend frequency} \\
    Degree of freedom & CPS to C-6D & C-6D to GS-13 \\
    \hline
    $X$ & 25\,mHz & 2\,Hz \\
    $Y$ & 25\,mHz & 2\,Hz \\
    $Z$ & \multicolumn{2}{c}{CPS to GS-13 - 100\,mHz} \\
    $RX$ & 7\,mHz & 2\,Hz \\
    $RY$ & 7\,mHz & 2\,Hz \\
    $RZ$ & 19\,mHz & 2\,Hz \\
    \hline
    \hline
    \end{tabular}
    \label{tab:6Dblend}
\end{table}

Active platform control at LIGO's MIT facility is based on the HAM-ISI control schemes discussed in Ref.~\cite{MatichardReview}. In this scheme, each degree of freedom is sensed and controlled independently as shown in Fig.~\ref{fig:MIT-loop}. 

In the longitudinal degrees of freedom, tilt-to-horizontal coupling~\cite{Tilt,Tilt2} appears as an effective noise source in the inertial sensing path. We augment the existing control scheme by introducing an additional inertial sensor (either BISs or C-6D) in the low-frequency sensing path. The improved tilt sensitivity suppresses the tilt-induced noise in the horizontal inertial readout. As a result, the sensor correction path to the capacitive position sensors is removed, eliminating the reinjection of ground tilt into the platform motion. In the case of C-6D, tilt and horizontal sensing are inherently coupled~\cite{6Dproto}; the decoupling scheme is discussed in more detail in the next section. We do not consider the high-frequency damping control ($>10$\,Hz) as our focus is on improved low-frequency suppression. The use of these technologies does not impact the high-frequency control, which can be implemented without change. 

We have demonstrated that both of our devices enable superior platform stabilization compared to the current scheme using GS-13s at the frequencies of interest below 1\,Hz. The devices do not replace existing seismic isolation infrastructure; instead, they are integrated in parallel, preserving the system robustness.

The blending frequencies for the three way blend between the CPS, C-6D or BISs, and the GS-13s implemented for these tests are shown in Tabs.~\ref{tab:BISblend} and~\ref{tab:6Dblend}.

\section{LIGO active platform simulations}\label{sec:platformSim}

We simulate two active platform configurations representative of those deployed at the LIGO facilities. We focus on the longitudinal ($X$) and pitch ($RY$) degrees of freedom, which couple directly to the interferometer sensitivity.

\begin{figure}[h]
    \centering
    \includegraphics[width=\linewidth]{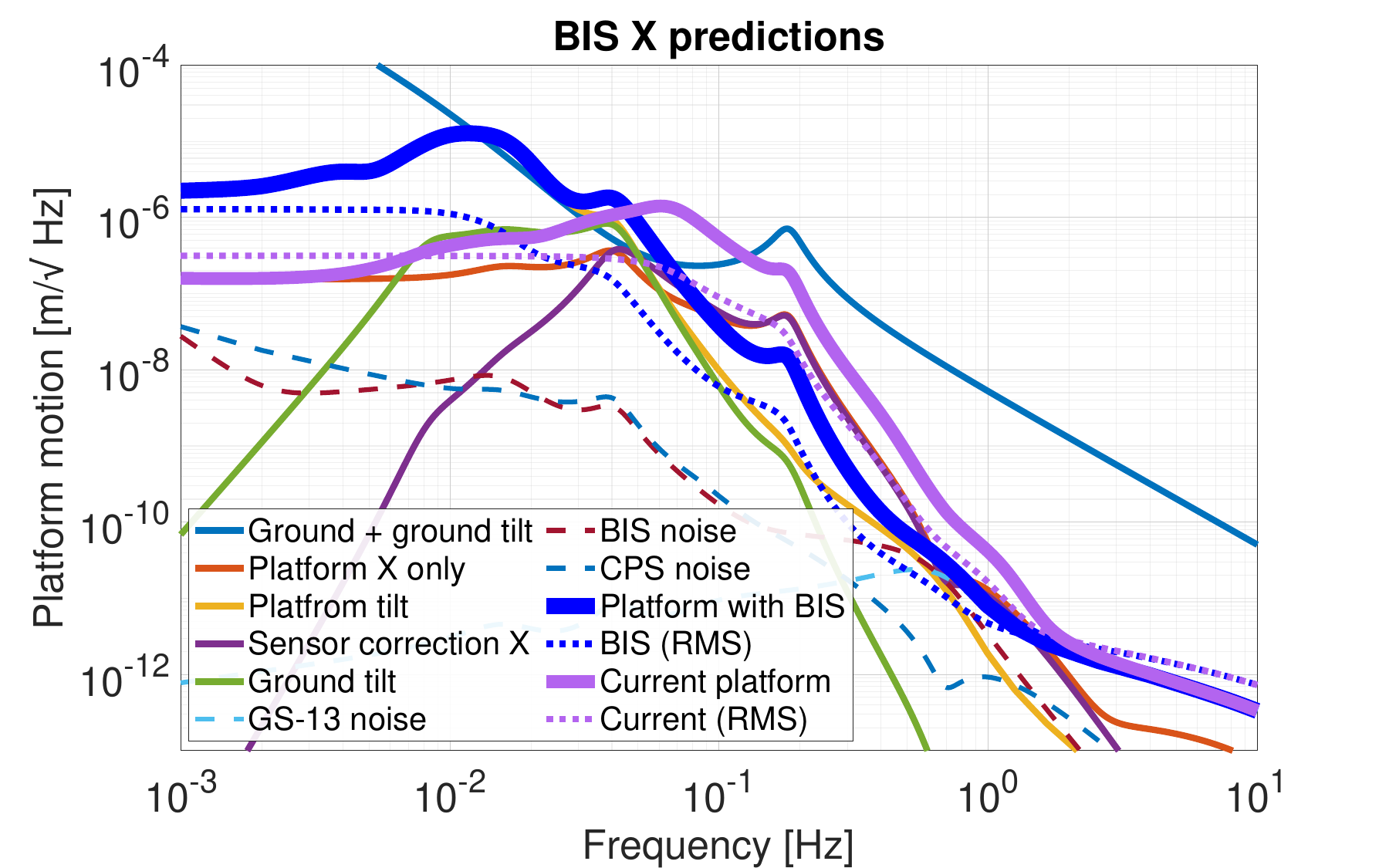}
    \caption{Simulated motion of the single stage platform when BISs are used in-loop. The dotted lines represent the RMS motion.}
    \label{fig:BISsim}
\end{figure}

\begin{table}[h]
    \caption{Comparison of blending frequencies for the single stage platform simulations.}
    \begin{tabular}{ccc}
    \hline
    \hline
     Degree of freedom & \multicolumn{2}{c}{Blend frequency} \\
    \hline
     & \multicolumn{2}{c}{CPS to GS-13}\\
    \hline
    $X$ & \multicolumn{2}{c}{100\,mHz} \\
    $RY$ & \multicolumn{2}{c}{250\,mHz} \\
    \hline
     & CPS to BIS & BIS to GS-13 \\
    \hline
    $X$ & 40\,mHz & 0.6\,Hz \\
    $RY$ & 23\,mHz & 0.6\,Hz \\
    \hline
    \hline
    \end{tabular}
    \label{tab:BISblendsim}
\end{table}

The first is a single-stage platform, identical to the system used at the MIT facility. Platform motion is modeled using the existing LIGO control infrastructure~\cite{MatichardReview}. The control scheme is illustrated in Fig.~\ref{fig:MIT-loop} for a single degree of freedom, but with the addition of CPS sensor correction.

In all configurations, tilt-to-horizontal coupling appears as an effective noise source in the inertial sensing path. This coupling limits horizontal isolation below $\sim 0.1$\,Hz for the current configuration, as shown by the violet trace in Fig.~\ref{fig:BISsim}.

We then simulate the addition of BIS sensors (green path in Fig.~\ref{fig:MIT-loop}) for low-frequency inertial control. The blending frequencies are shown in Tab.~\ref{tab:BISblendsim}. The improved tilt sensitivity suppresses platform tilt, thereby reducing horizontal motion above 50\,mHz relative to the baseline configuration (blue trace in Fig.~\ref{fig:BISsim}). At the microseismic peak, this results in approximately an order-of-magnitude improvement. Below 50\,mHz, residual platform tilt limits performance due to the lower blending frequencies. This trade-off increases low-frequency motion but reduces RMS motion near the suspension resonances, improving operational robustness. 

\begin{figure}[h]
\centering
    \includegraphics[width = \linewidth]{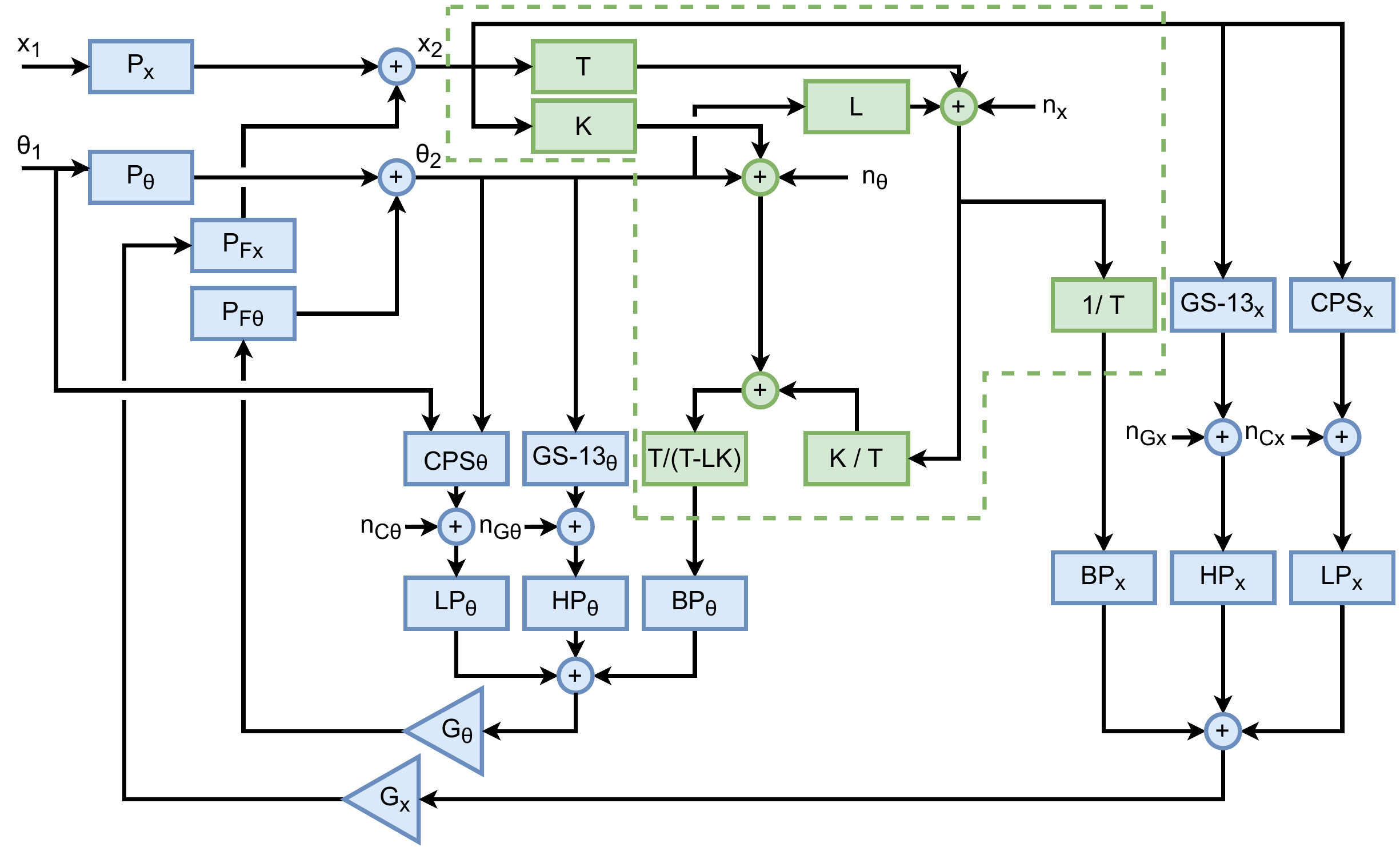}
    \caption{Control diagram when integrating C-6D for active platform control. The decoupling scheme for longitudinal ($X$) and pitch ($RY$) are shown in light green. $T$ is the translation-to-translation sensing transfer function, $K$ is translation-to-tilt, and $L$ is the fiber length for C-6D.}
    \label{fig:control_diagram}
\end{figure}

\begin{figure}
    \centering
    \includegraphics[width=\linewidth]{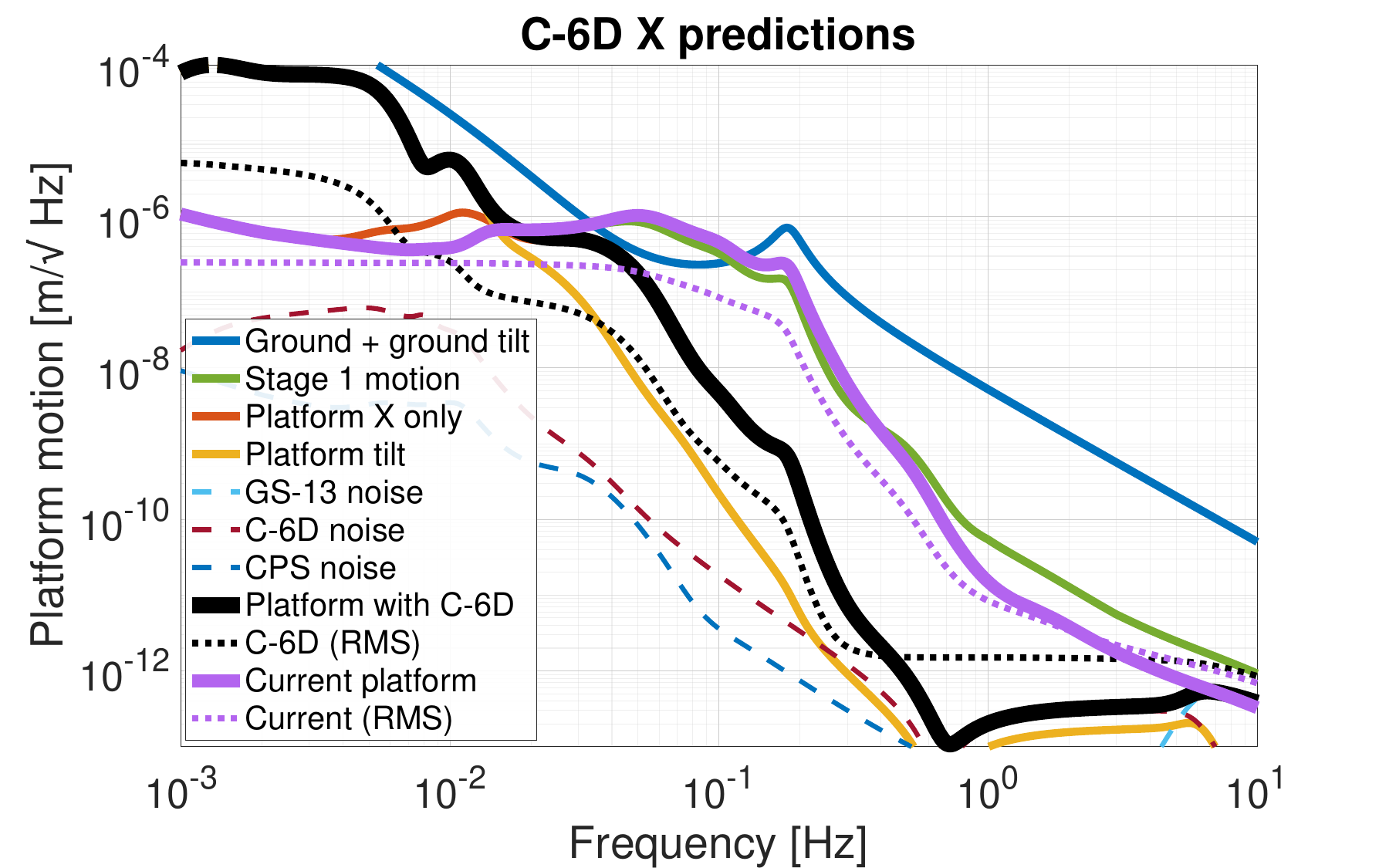}
    \caption{Simulated motion of the two-stage platform when C-6D is used in-loop. Dotted lines represent the RMS motion.}
    \label{fig:6Dsim}
\end{figure}

\begin{table}[h]
    \caption{Comparison of blending frequencies for two-stage platform simulations.}
    \begin{tabular}{ccc}
    \hline
    \hline
    Degree of freedom & \multicolumn{2}{c}{Blend frequency} \\
    \hline
    Stage 1 & \multicolumn{2}{c}{CPS to T-240} \\
    \hline
    $X$ & \multicolumn{2}{c}{100\,mHz} \\
    $RY$ & \multicolumn{2}{c}{170\,mHz} \\
    \hline
    Stage 2 & \multicolumn{2}{c}{CPS to GS-13} \\
    \hline
    $X$ & \multicolumn{2}{c}{240\,mHz} \\
    $RY$ & \multicolumn{2}{c}{stage 1 $RY$ = stage 2 $RY$} \\
    \hline
    Stage 2 & CPS to C-6D & C-6D to GS-13 \\
    \hline
    $X$ & 11\,mHz & 6\,Hz \\
    $RY$ & 3\,mHz & 6\,Hz \\
    \hline
    \hline
    \end{tabular}
    \label{tab:6Dblendsim}
\end{table}

We also simulate a two-stage active platform configuration used to suspend the LIGO test mass optics~\cite{MatichardPart1,MatichardPart2,MatichardReview}. In this architecture, the first stage follows a similar scheme to Fig.~\ref{fig:MIT-loop} with CPS sensor correction, but utilizes T-240 seismometers for inertial sensing. The second stage employs GS-13 seismometers and does not include sensor correction. As in the single-stage system, each degree of freedom is sensed and controlled independently, with tilt-to-horizontal coupling entering the longitudinal inertial readout as an effective noise source. The blending frequencies are shown in Tab.~\ref{tab:6Dblendsim}, and the resulting stage-two platform motion is shown in violet in Fig.~\ref{fig:6Dsim}. 

We evaluate the impact of integrating C-6D on the second stage. Because C-6D inherently couples horizontal and tilt sensing, a decoupling scheme is required, as described in Ref.~\cite{6Dplatform} and illustrated in green in Fig.~\ref{fig:control_diagram}. We decouple only horizontal motion from the tilt readout to avoid injecting tilt-sensing noise into the horizontal control path. The tilt-to-horizontal coupling therefore remains in the horizontal readout, but the decoupling enables independent control of all six degrees of freedom.

The enhanced tilt sensitivity suppresses residual ground tilt by approximately an order of magnitude at 10\,mHz. This leads to improved horizontal platform isolation between $15\,\rm{mHz} - 6\,\rm{Hz}$, with up to two orders-of-magnitude additional suppression at the microseismic peak. As in the single-stage case, lowering the blend frequencies introduces additional low-frequency tilt. The additional low-frequency motion remains below typical Earth tide amplitudes and does not compromise interferometer stability. However, the net effect is a substantial reduction in RMS motion near the suspension resonances, improving lock acquisition robustness, detector stability, and noise performance.

\section{LIGO interferometer modeling}

\begin{figure}
    \centering
    \includegraphics[width=1\linewidth]{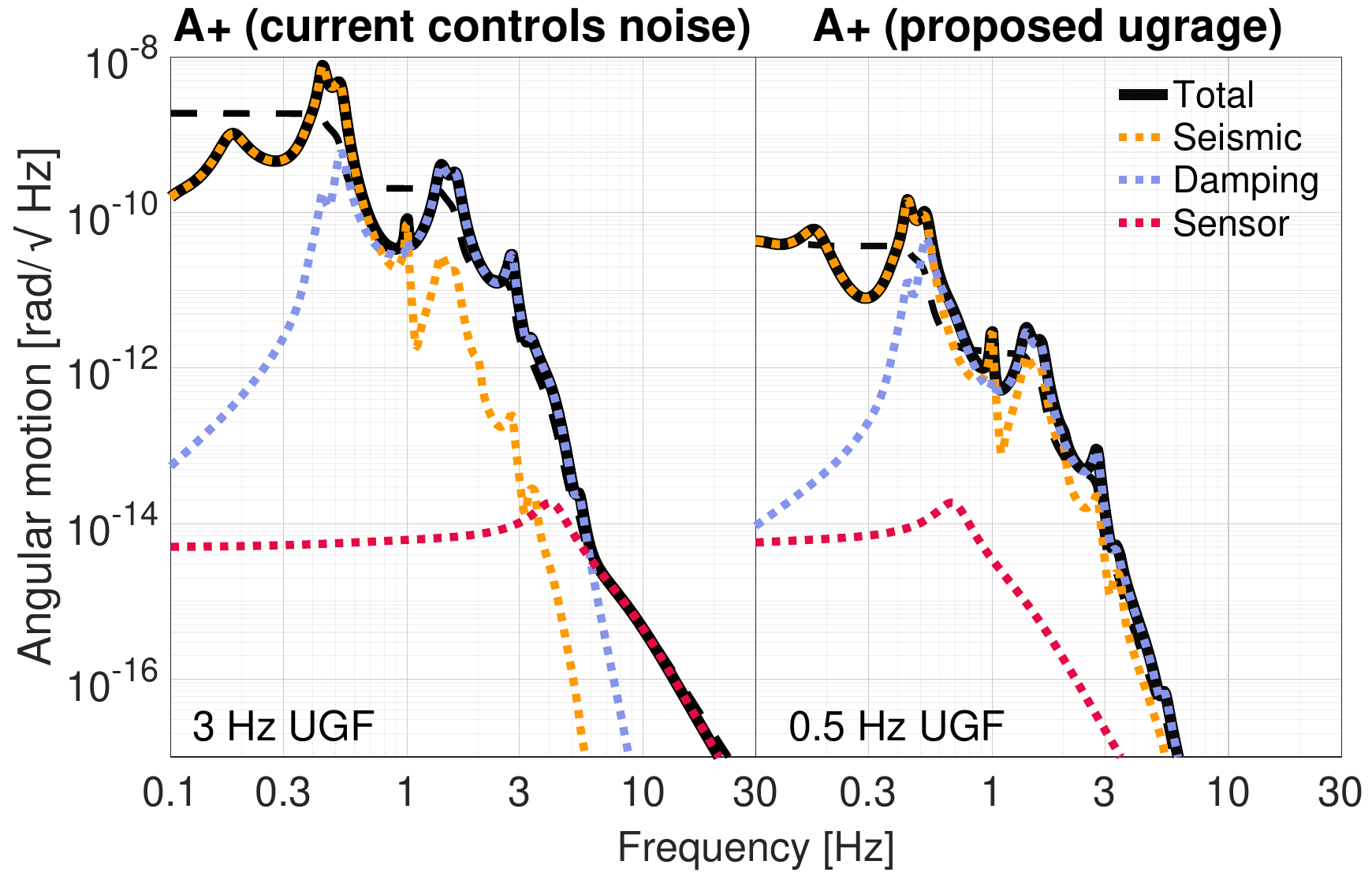}
    \caption{Comparison of the test mass pitch motion, showing the contributions of residual seismic motion, suspension damping noise, and global control sensor noise. The transfer functions of the QUAD suspension model were used to generate this figure.}
    \label{fig:TMpitch}
\end{figure}

The sensitivity of the interferometer is determined by the residual motion of the suspended optics. Two key control noises dominate the strain sensitivity below 30\,Hz \cite{O4sensitivity, O4characterisation}. The first is alignment sensing and control (ASC), which is dominated by test-mass pitch motion, $\theta_{TM}$. We propagate the simulated two-stage platform motion through the quadruple suspension transfer functions to obtain the residual opti motion. We compare the current control scheme, which uses commercial seismometers and optical shadow sensors, with the proposed upgrade configuration, which uses C-6D and LPSs. A global control scheme using wavefront sensors is implemented to further reduce the motion for both configurations. In the current configuration, the global ASC bandwidth must be maintained at approximately 3\,Hz~\cite{DetCharISC} to suppress RMS angular motion. However, this bandwidth injects sensor noise into the gravitational-wave band. In the proposed upgrade configuration, improved seismic isolation and suspension damping allow the ASC bandwidth to be reduced to 0.5\,Hz while simultaneously decreasing the RMS motion by a factor of 10 (Fig.\,\ref{fig:TMpitch}). The improved seismic isolation reduces the required global control authority, thereby decreasing the injection of global sensor noise into the strain channel. We assume that the motion of the four test masses is uncorrelated. We also assume a beam offset of $d_{oc} = 1$\,mm. The beam offset introduces a direct angle-to-length coupling. The residual angular pitch motion is used as a proxy for the total angular motion of the test masses. We assume the total angular motion (pitch and yaw) couples to the gravitational-wave strain as $h_{ASC} = 2\sqrt{2} d_{oc}/L_{arm} \times \theta_{TM} $, where $L_{arm} = 4$\,km is the length of the arm cavities. 

\begin{figure}
    \centering
    \includegraphics[width=\linewidth]{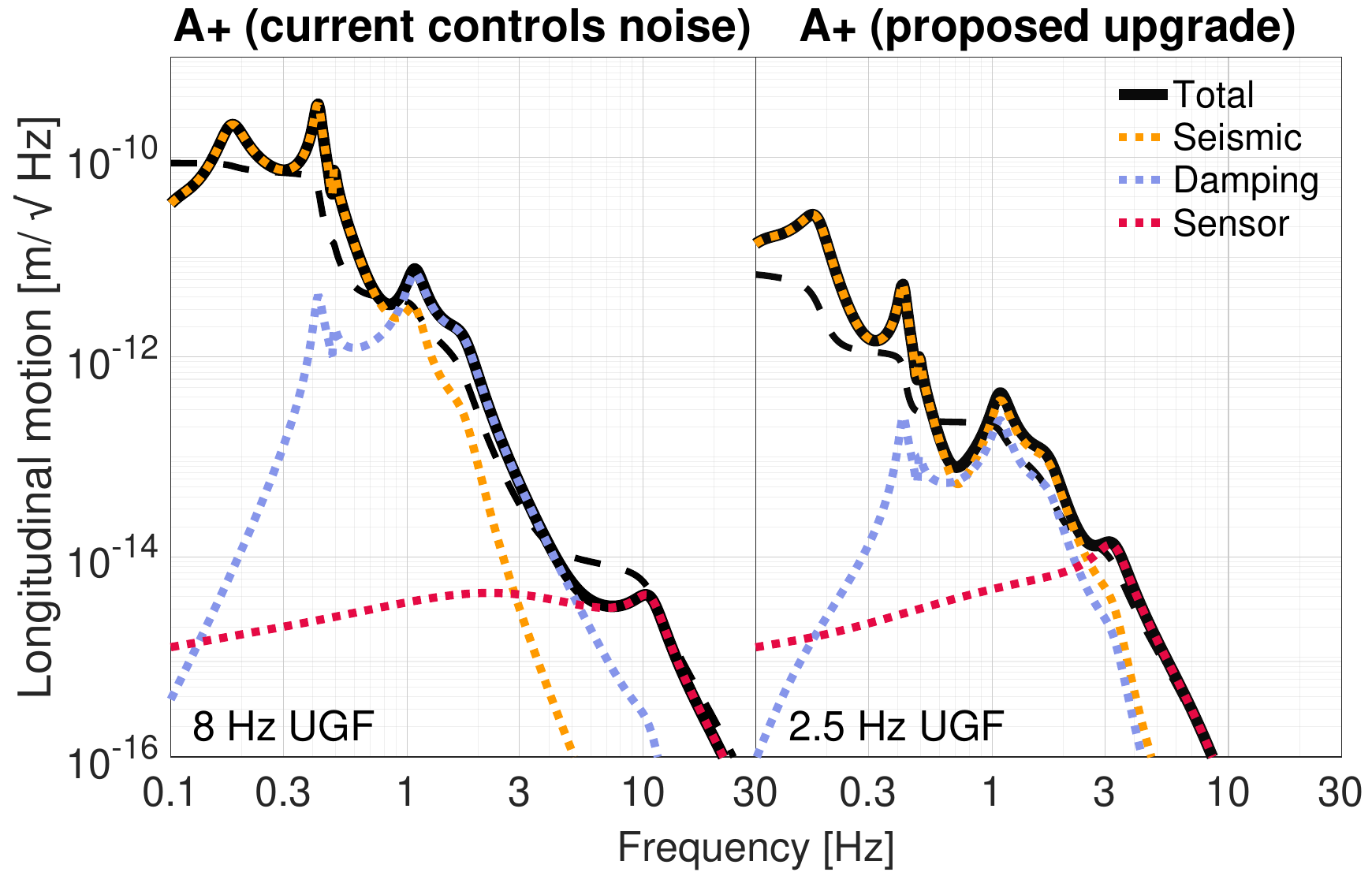}
    \caption{Comparison of the longitudinal beam splitter motion, showing the contributions of residual seismic motion, suspension damping noise, and global control sensor noise. The transfer functions of the BSFM triple suspension model were used to generate this figure.}
    \label{fig:BSlong}
\end{figure}

The second dominant control noise is length sensing and control, primarily the Michelson degree of freedom (MICH). This degree of freedom is dominated by longitudinal beam splitter motion. We shape the global sensor noise to match the measured excess noise relative to the noise budget shown in Fig.\,9(a) of Ref.~\cite{O1sensitivity}. The proposed upgrade enables the control bandwidth to be reduced from 11\,Hz~\cite{DetCharISC} to 8\,Hz (Fig.~\ref{fig:BSlong}). The RMS motion is simultaneously reduced by a factor $>10$. The beamsplitter motion, $x_{BS}$, couples to the gravitational-wave strain as $h_{MICH} = \sqrt{2} / G_{arm} \times x_{BS}/L_{arm}$, where $G_{arm} = 260$ is the arm-cavity amplification factor. An additional low-pass filter is applied to simulate the feedforward control scheme to reduce the MICH noise, achieving a contribution to the strain noise similar to the measured noise in Fig.\,(5) in Ref.~\cite{O4sensitivity}.

\begin{figure}
    \centering
    \includegraphics[width=1\linewidth]{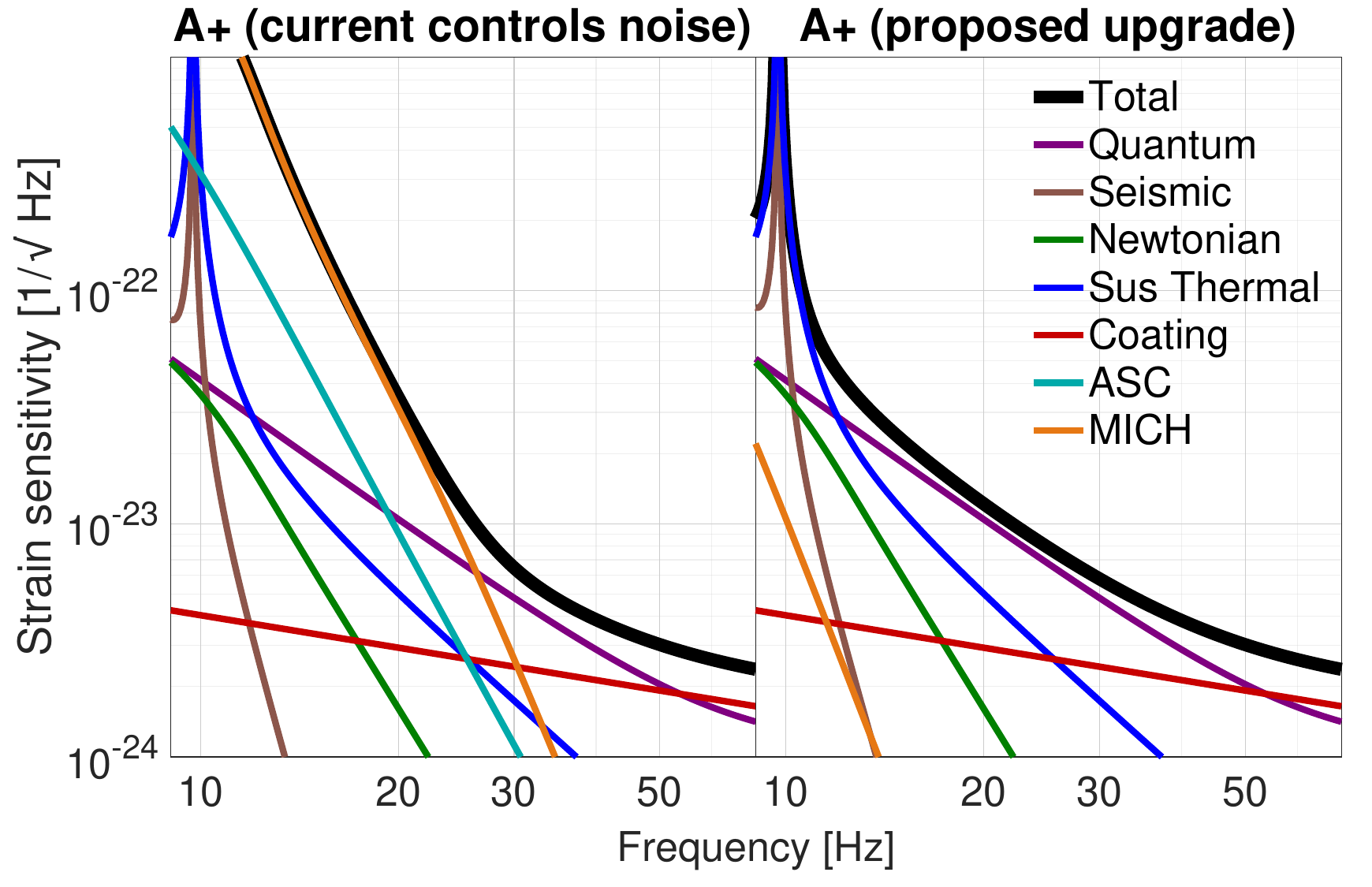}
    \caption{Strain noise budgets produced using pyGWINC~\cite{pygwinc} with the inclusion of ASC and MICH noise. A single-pole low-pass filter was applied to the MICH noise at 2\,Hz to simulate the feedforward scheme. Subdominant contributions to the total noise have been omitted from the plots for clarity.}
    \label{fig:StrainNB}
\end{figure}

For the modeled proposed upgrade configuration, both ASC and MICH control noise contributions fall below the projected fundamental detector noise. The strain noise budgets were generated using pyGWINC~\cite{pygwinc} and are shown in Fig.~\ref{fig:StrainNB}.

\begin{figure}[h]
    \centering
    \includegraphics[width=\linewidth]{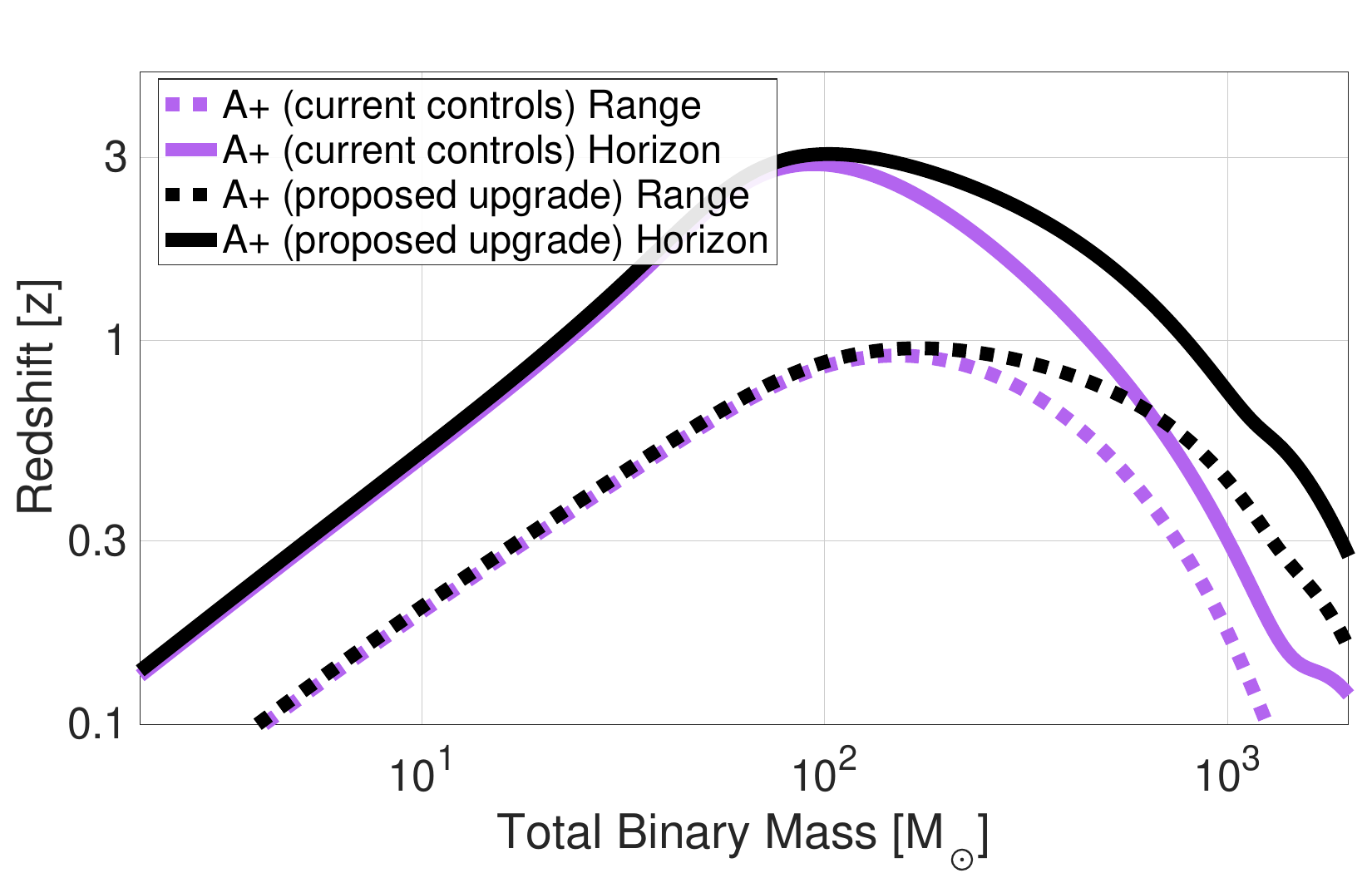}
    \caption{Horizon and range assuming equal mass binaries. The violet trace represents the A+ design sensitivity with the addition of current technical noise. The black trace is with the implementation of our technologies in the A+ detector configuration.}
    \label{fig:HR}
\end{figure}

To quantify the astrophysical impact, the strain budgets were used to compute the detect horizon and range using the inspiral-range calculator~\cite{inspiralrange} assuming equal mass binaries. The results are illustrated in Fig.~\ref{fig:HR}, demonstrating a factor 3 improvement for $1000\,M_{\odot}$ sources.

Furthermore, we simulate improvements to the auxiliary controls. We focus on the signal-recycling cavity due to its direct impact on recovering the gravitational-wave signal. The angular motion of the telescope mirror determines the signal recycling cavity axis. The axis differs for the radio-frequency sidebands used for interferometer control and the carrier light that contains the gravitational-wave signal. Because the interferometer operates at the dark fringe, there is a large disparity in optical power between these fields. The radio-frequency sidebands carry 10–100 times more power than the carrier. This power imbalance complicates simultaneous stabilization of the cavity axis for both control and signal fields, making it necessary to reduce the residual motion of telescope mirror without relying on additional global control loops. 

\begin{figure}
    \centering
    \includegraphics[width=1\linewidth]{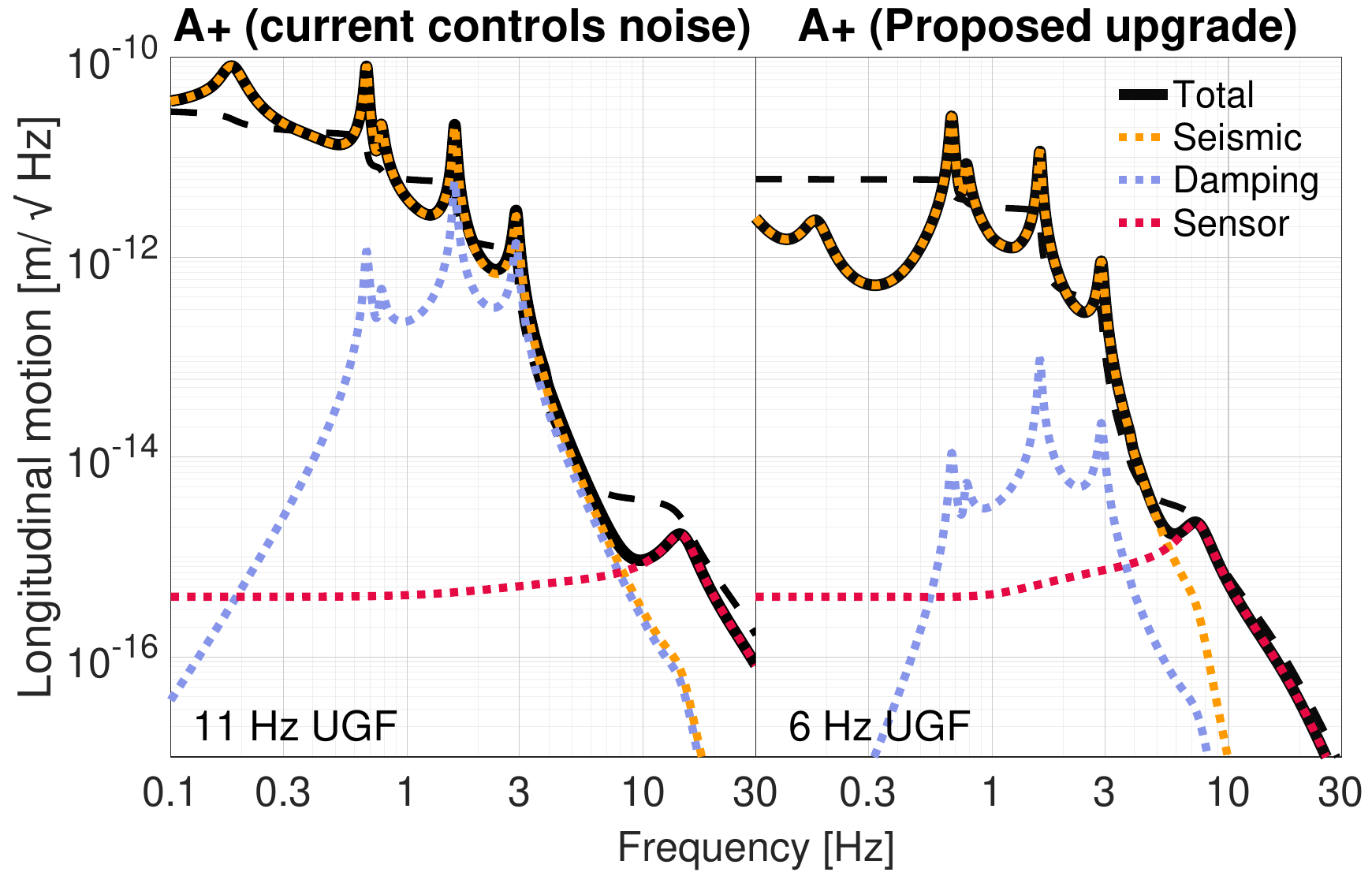}
    \caption{Comparison of the signal-recycling telescope optic longitudinal motion, showing the contributions of residual seismic motion, suspension damping noise, and global control sensor noise. The transfer functions of the HAM Large Triple Suspension (HLTS) model were used to generate this figure.}
    \label{fig:SR3long}
\end{figure}

\begin{figure}
    \centering
    \includegraphics[width=1\linewidth]{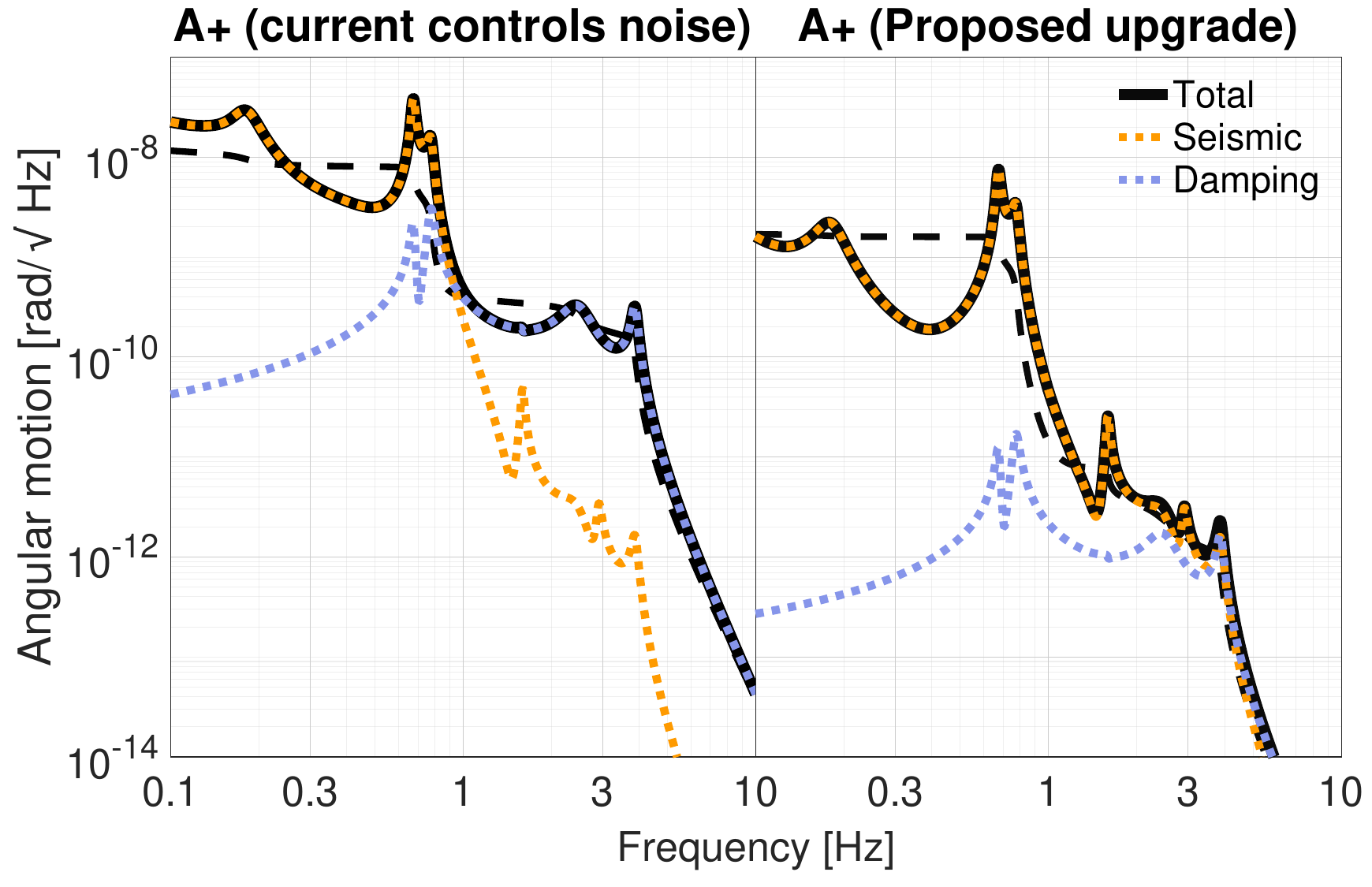}
    \caption{Comparison of the signal-recycling telescope optic pitch motion, showing the contributions of residual seismic motion, and suspension damping noise. No global control is assumed here. The transfer functions of the HLTS model were used to generate this figure.}
    \label{fig:SR3pitch}
\end{figure}

We model the residual motion of the telescope optic in a similar fashion to the test masses and beam splitter. The single-stage platform motion is propagated through the HAM Large Triple Suspension model. This enables us to compare the use of the BISs and LPSs (proposed upgrade) compared to the current configuration (current controls noise). The residual longitudinal and pitch motion is shown in Fig.~\ref{fig:SR3long} and Fig.~\ref{fig:SR3pitch} respectively. The angular motion RMS is reduced by a factor of 5 for the proposed upgrade configuration. This reduces output beam pointing fluctuations that couple through the output mode cleaner. This is expected to improve calibration-line stability, reduce control-noise constraints due to pointing to DC intensity noise coupling, and increase robustness of the detector to elevated microseismic motion.

\bibliographystyle{apsrev4-2}  
\bibliography{main.bib}